%% file: main.tex
\documentclass[sigconf, screen]{acmart}
\usepackage{balance}
\usepackage{microtype}

\usepackage{amsmath,amsfonts}

\usepackage[]{algorithm}
\newfloat{algorithm}{t}{lop}
\usepackage[noend]{algpseudocode}

\usepackage{balance}
\usepackage{textcomp}
\usepackage{xspace}
\usepackage{xcolor}
\definecolor{annotatecolor}{rgb}{0.59,0,0.09}
\usepackage{graphicx}
\usepackage[normalem]{ulem} %
\usepackage{enumitem}

\usepackage{hyperref}
\hypersetup{linkcolor=black,citecolor=black,anchorcolor=black,filecolor=black,menucolor=black,runcolor=black,urlcolor=black,hidelinks}
\usepackage{breakurl}

\usepackage{booktabs}
\usepackage{multirow}
\usepackage{makecell}
\usepackage{ragged2e}

\usepackage{caption}
\usepackage{subcaption}
\usepackage{subfloat}

\usepackage{listings}

\definecolor{gray}{RGB}{211,211,211}
\newcommand{\jbasicstyle}{\small\ttfamily} %

\newcommand{\jnumberstyle}{\scriptsize}

\lstdefinelanguage{pseudo}
{
morekeywords={},
keywordstyle=\bfseries,
lineskip=-0.1em,
numbers=left, %
numberstyle=\jnumberstyle,
numbersep=4pt,
basicstyle=\jbasicstyle,
breaklines=true,
breakautoindent=true,
tabsize=2,
columns=fullflexible,
morecomment=*[l][\textsl]{//},
mathescape=true,
xleftmargin=10pt,
}

\lstdefinelanguage{todo-comment}
{
morekeywords={},
keywordstyle=\bfseries,
lineskip=-0.1em,
numbers=none,
basicstyle=\scriptsize\ttfamily,
breaklines=true,
breakautoindent=true,
tabsize=2,
columns=fullflexible,
morecomment=*[l][\textsl]{//},
mathescape=true,
xleftmargin=0pt,
}

\definecolor{keywordcolor}{rgb}{0,0,1}      %
\definecolor{modifiercolor}{rgb}{0.5,0,0.5} %
\definecolor{datatypecolor}{rgb}{0.82,0.16,0.46} %
\definecolor{methodcolor}{rgb}{0.25,0.5,0.35} %
\definecolor{byzantine}{rgb}{0.74, 0.2, 0.64}  %
\definecolor{cadetblue}{rgb}{0.37, 0.62, 0.63}  %
\definecolor{cadet}{rgb}{0.0, 0.42, 0.24}
\definecolor{brown(web)}{rgb}{0.65, 0.16, 0.16}  %
\definecolor{bluegray}{rgb}{0.2, 0.2, 0.6}

\lstdefinelanguage{java-pretty}
{
language=java,
numbers=left,
basicstyle=\scriptsize\ttfamily,
numberstyle=\scriptsize,
breaklines=true,
columns=fullflexible,
xleftmargin=14pt,
tabsize=2,
showstringspaces=false,
deletekeywords={public, private, protected, static, final, class, interface, abstract, implements, extends, if, else, while, do, for, switch, case, default, break, continue, return, int, long, double, float, boolean, char, void, String,this},
morekeywords=[1]{if, else, while, do, for, switch, case, default, break, continue, return},
keywordstyle=[1]\color{byzantine}\bfseries,
morekeywords=[2]{public, private, protected, static, final, class, interface, abstract, implements, extends},
keywordstyle=[2]\color{bluegray}\bfseries,
morekeywords=[3]{int, long, double, float, boolean, char, void, String, @Override, @Test},
keywordstyle=[3]\color{cadet}\bfseries,
morekeywords=[4]{class, interface, extends, implements, new, super, throw, throws, try, catch, finally},
keywordstyle=[4]\color{methodcolor},
morecomment=[l]{//},
commentstyle=\color{cadet},
stringstyle=\color{brown(web)},
}

\usepackage{tikz}
\usetikzlibrary{calc}
\usetikzlibrary{shapes.geometric}
\usetikzlibrary{decorations.pathreplacing}
\usetikzlibrary{positioning}

\usepackage{datetime} %

\definecolor{darkgreen}{rgb}{0.0, 0.5, 0.0}
\newcommand{\XSpace}[1]{}
\newcommand{\XComment}[1]{}
\newcommand{\Fix}[1]{\textcolor{red}{#1}}

\newcommand{\EditRm}[1]{}

\newcommand{\DefMacro}[2]{\expandafter\newcommand\csname rmk-#1\endcsname{#2}}
\newcommand{\UseMacro}[1]{\csname rmk-#1\endcsname}

\newcommand{\MyPara}[1]{\noindent\textbf{#1}.}

\newcommand{\InputWithSpace}[1]{\bgroup\def\arraystretch{1.1}\input{#1}\egroup}
\newcommand{\Code}[1]{{\ifmmode{\mathtt{#1}}\else$\mathtt{#1}$\fi}}
\newcommand{\CodeIn}[1]{{\ifmmode{\mathtt{#1}}\else$\mathtt{#1}$\fi}}

\newcolumntype{R}[1]{>{\RaggedLeft\arraybackslash}p{#1}}
\newcolumntype{L}[1]{>{\RaggedRight\arraybackslash}p{#1}}

\newcommand{\ltrue}{\top} %
\newcommand{\lfalse}{\bot} %

\newcommand{\Tool}{\textsc{exLong}\xspace}

\newcommand{\CAT}{CAT-LM\xspace}
\newcommand{\GPTThree}{GPT3.5\xspace}
\newcommand{\GPTFour}{GPT-4o\xspace}
\newcommand{\exlongGPTFour}{\Tool--GPT-4o\xspace}

\newcommand{\Java}{Java\xspace}

\newcommand{\TeCo}{TeCo\xspace}
\newcommand{\Title}{A Tool for Generating Exceptional Behavior Tests With Large Language Models}

\newcommand{\ollama}{ollama\xspace}
\newcommand{\python}{Python\xspace}
\newcommand{\eg}{e.g.\xspace}
\newcommand{\ie}{i.e.\xspace}
\newcommand{\userView}{developer-oriented use case\xspace}
\newcommand{\UserView}{Developer-oriented use case\xspace}

\newcommand{\machineView}{machine-oriented use case\xspace}

\newcommand{\MachineView}{Machine-oriented use case\xspace}

\newcommand{\EBTs}{EBTs\xspace}
\newcommand{\EBT}{EBT\xspace}

\newcommand{\nEBTs}{non-EBTs\xspace}
\newcommand{\nEBT}{non-EBT\xspace}

\newcommand{\MUT}{MUT\xspace}

\newcommand{\LLM}{LLM\xspace}

\newcommand{\stacktrace}{stack trace\xspace}
\newcommand{\stacktraces}{stack traces\xspace}

\newcommand{\guardexp}{guard expression\xspace}
\newcommand{\guardexps}{guard expressions\xspace}

\newcommand{\testfile}{destination test file\xspace}

\newcommand{\ts}{target throw statement\xspace}

\newcommand{\tss}{target throw statements\xspace}

\newcommand{\gt}{ground-truth\xspace}
\newcommand{\CodeLlama}{CodeLlama\xspace}

\newcommand{\finetuned}{fine-tuned\xspace}

\newcommand{\repo}{repository\xspace}

\newcommand{\Randoop}{Randoop\xspace}
\newcommand{\EvoSuite}{EvoSuite\xspace}

\newcommand{\etype}{etype\xspace}
\newcommand{\aTest}{\ensuremath{\mathtt{t}}\xspace}
\newcommand{\aETest}{\ensuremath{\aTest_{\mathtt{eb}}}\xspace}
\newcommand{\aNETest}{\ensuremath{\aTest_{\mathtt{neb}}}\xspace}

\newcommand{\aPrompt}{\ensuremath{\mathtt{p}}\xspace}

\newcommand{\aThrows}{\ensuremath{\mathtt{E}}\xspace}
\newcommand{\sTrace}{\ensuremath{STrace}\xspace}

\newcommand{\aThrowsCovered}{\ensuremath{\aThrows^{\ltrue}}\xspace}
\newcommand{\aThrowsUncovered}{\ensuremath{\aThrows^{\lfalse}}\xspace}
\newcommand*\circled[1]{\tikz[baseline=(char.base)]{
\node[shape=circle, draw=black, minimum size=0.1, inner sep=0.8pt, fill=white, thick, text=black] (char) {#1};}}

\DefMacro{TCap-rq1-eval-dataset-stats}{Statistics of the evaluation dataset for \userView.\xspace\#\MUT is the number of unique method under test; \#Exception Types is the number of unique exception types tested by the \EBTs; \#Throw Statements is the number of unique throw statements covered by the \EBTs.\vspace{-5pt}}
\DefMacro{TCap-analysis-based-tools}{Randoop and EvoSuite generated tests statistics for the
30 projects we studied. We report number of tests generated by each tool and the time taken for generation per project. 0 means the tool fails to execute on that project. We report the number of throw statements that are within the public method and number of throw statements we used for evaluating tools for each project.\vspace{-5pt}}
\DefMacro{TCap-dataset}{Statistics of the collected dataset. \#MUTs is the number of unique method under test; \#Exception Types is the number of unique exception types tested by the \EBTs.\vspace{-5pt}}
\DefMacro{TCap-eval-dataset-stats}{Statistics of the evaluation dataset for \machineView. \#Throw Statements is the number of \tss we extracted from the repository according to Section~\ref{sec:tech:eval:usecases}.\xspace\#Exception Types is the number of unique exception types thrown by the \tss.
\vspace{-5pt}}
\DefMacro{TCap-exlong-runtime}{Runtime cost of \Tool for each project.}
\DefMacro{TCap-Tool-cmp-table}{Throw statements coverage rate for \Tool, \Randoop and \EvoSuite.\vspace{-5pt}}%
\DefMacro{TCap-models-model-size}{Results for \Tool-base 13B and \Tool 13B models; \Tool-base 13B does not have access to \nEBTs, \stacktrace and \guardexp.\vspace{-5pt}}
\DefMacro{TCap-models-user-view-with-name}{Results on \userView with \gt \EBT's name in the prompt.\vspace{-5pt}}
\DefMacro{TCap-models-user-view-no-name}{Results on \userView without \gt \EBT's name in the prompt.\vspace{-5pt}}
\DefMacro{TCap-models-exlong-ablation}{Ablations on different context of \Tool.\vspace{-5pt}}
\DefMacro{TCap-models-netest-diversity}{Comparison between using different \nEBTs to sample and using the same \nEBT but sampling multiple times. Note that we add the target \EBT's name to the prompt and we only report results on examples that have more than one candidate \nEBT.\vspace{-5pt}}
\DefMacro{TCap-mut2e-pattern}{Number of instances for each \EBT pattern~\cite{MarcilioFuria21HowJavaProgrammersTestExceptionalBehavior} in the collected corpus.\vspace{-5pt}}
\DefMacro{TCap-models-13b-and-7b}{Comparison between 7B and 13B results.\vspace{-5pt}}
\DefMacro{TCap-models-gpt4o-with-name}{\MR{Comparison between \GPTFour-few-shot and \exlongGPTFour.}}
\DefMacro{TH-num-project}{\#project\xspace}
\DefMacro{TH-num-module}{\#module\xspace}
\DefMacro{TH-num-test}{\#\aTest\xspace}
\DefMacro{TH-num-etest-cover-tc}{\# ET-tc \xspace}
\DefMacro{TH-num-etest-undeclared-exception}{\# ET udex\xspace}
\DefMacro{TH-num-etest-unk-coverage}{\# tests nomut\xspace}
\DefMacro{TH-num-etest}{\#\aETest\xspace}
\DefMacro{TH-num-throws}{\#\aThrows\xspace}
\DefMacro{TH-num-covered}{\#\aThrowsCovered\xspace}
\DefMacro{TH-num-self-thrown}{\#SelfThrown\xspace}
\DefMacro{TH-coverage}{coverage[\%]\xspace}
\DefMacro{TH-num-ne-e-pair}{\#$(\aNETest,\aETest)$\xspace}
\DefMacro{TH-num-uncovered-w-ne}{\#\aThrowsUncovered w/ \aNETest\xspace}
\DefMacro{TH-source-m}{Manual\xspace}
\DefMacro{TH-source-r}{\Randoop\xspace}
\DefMacro{TH-source-u}{Union\xspace}

\DefMacro{TH-num-etest-not-runnable}{\#no-run\xspace}
\DefMacro{TH-num-etest-covered-throw-clause}{\# check\xspace}
\DefMacro{TH-num-etest-unchecked-exception}{\# uncheck \xspace}
\DefMacro{TH-num-etest-unmatched-exception}{\# unma \xspace}
\DefMacro{TH-num-etest-no-mut}{\#no mut\xspace}
\DefMacro{TH-num-etest-no-throw-stmt}{\# no thr \xspace}
\DefMacro{TH-num-w-condition}{\# with conditions\xspace}

\DefMacro{THead-subset-projects}{Subset Projects}
\DefMacro{THead-all-projects}{All Projects}

\DefMacro{THead-ebts}{\#\EBTs\xspace}
\DefMacro{THead-count-mut}{\#\MUT\xspace}
\DefMacro{THead-rq1-eval}{Developer-Oriented}
\DefMacro{THead-models}{Models\xspace}
\DefMacro{THead-coverage-max}{ThrowCov\%}
\DefMacro{THead-bleu-avg}{BLEU\xspace}
\DefMacro{THead-bleu-max}{BLEU\xspace}
\DefMacro{THead-code-bleu-avg}{CodeBLEU\xspace}
\DefMacro{THead-code-bleu-max}{CodeBLEU\xspace}
\DefMacro{THead-edit-sim-avg}{EditSim\xspace}
\DefMacro{THead-edit-sim-max}{EditSim\xspace}
\DefMacro{THead-xmatch-top1}{xMatch\xspace}
\DefMacro{THead-compilable-avg}{Compilable\%\xspace}
\DefMacro{THead-compilable-max}{Compilable\%\xspace}
\DefMacro{THead-match-avg}{Matched-E\%\xspace}
\DefMacro{THead-match-max}{Matched-E\%\xspace}
\DefMacro{THead-runnable-avg}{Runnable\%\xspace}
\DefMacro{THead-runnable-max}{Runnable\%\xspace}
\DefMacro{THead-runnable-overall-max}{Runnable\%\xspace}
\DefMacro{THead-train}{Train\xspace}
\DefMacro{THead-test}{Eval\xspace}
\DefMacro{THead-valid}{Valid\xspace}
\DefMacro{THead-project-count}{\#Projects\xspace}
\DefMacro{THead-count-project}{\#Projects\xspace}
\DefMacro{THead-with-stack-trace-count}{With \sTrace\xspace}
\DefMacro{THead-max-stack-trace-len}{Max. len \sTrace\xspace}
\DefMacro{THead-median-stack-trace-len}{Med. len \sTrace\xspace}
\DefMacro{THead-min-stack-trace-len}{Min. len \sTrace\xspace}
\DefMacro{THead-median-stack-trace-count}{Med. \# \sTrace\xspace}
\DefMacro{THead-max-stack-trace-count}{Max. \# \sTrace\xspace}
\DefMacro{THead-with-netest-count}{With \aNETest\xspace}
\DefMacro{THead-max-netest-count}{Max. \# \aNETest\xspace}
\DefMacro{THead-median-netest-count}{Med. \# \aNETest\xspace}
\DefMacro{THead-exception-count}{\# \etype\xspace}
\DefMacro{THead-count-eval-throw-stmt-data}{\# \aETest \xspace}
\DefMacro{THead-count-projects}{\#Projects}
\DefMacro{THead-count-module}{\#Modules}
\DefMacro{THead-count-etype}{\#Ex. Type\xspace}
\DefMacro{THead-count-ts}{\#ThrowSt}
\DefMacro{THead-rq1}{User View}
\DefMacro{THead-rq2}{Machine View}

\DefMacro{THead-count-static-analysis-helps}{\# Analysis find\xspace}
\DefMacro{THead-count-methods-with-ts}{\# M. w. ts}
\DefMacro{THead-count-methods-with-ts-public}{\# Public M}

\DefMacro{THead-count-ts-public}{\# Pub. Ts.}
\DefMacro{THead-mean-ts-per-method}{Avg. \# TS}
\DefMacro{THead-coverage}{Exception coverage}
\DefMacro{THead-exception-coverage-evosuite}{\EvoSuite Coverage}
\DefMacro{THead-exception-coverage-randoop}{\Randoop Coverage}

\DefMacro{TCap-eval-dataset-ts-stats}{Stats on Methods from testset projects with Throw Statements. \# methods with throw statements, \# non-private and non-protected methods among them, \# total throw statements, Avg. \# throw statements per method.}
\DefMacro{THead-selected-253-conditionnestack2e-with-name-ft-lora-codellama-7b}{\Tool}
\DefMacro{THead-selected-253-conditionnestack2e-all-with-name-ft-lora-codellama-7b}{\Fix{\ToolMoreNebt w/ different \nEBTs}} %
\DefMacro{THead-conditionnestack2e-sample-with-name-ft-lora-codellama-7b}{\ToolMoreNebt w/ same \nEBT} %
\DefMacro{THead-diversity-conditionnestack2e-all-with-name-ft-lora-codellama-7b}{\ToolMoreNebt w/ different \nEBT}
\DefMacro{THead-selected-434-mut2e-no-name-ft-lora-codellama-7b}{MUT+FILE2E\xspace}
\DefMacro{THead-selected-434-mut2e-with-name-ft-lora-codellama-7b}{No \stacktrace, \guardexp, \nEBT}
\DefMacro{THead-mut2e-with-name-ft-lora-codellama-7b}{No \stacktrace, \guardexp, \nEBT}
\DefMacro{THead-selected-434-mut2e-with-name-ft-lora-codellama-13b}{\Tool-base 13B\xspace}
\DefMacro{THead-selected-434-mut2e-no-name-ft-lora-codellama-13b}{MUT+FILE2E(13B)\xspace}
\DefMacro{THead-selected-434-catlm-mut2e-with-name-catlm}{CAT-LM\xspace}
\DefMacro{THead-selected-434-catlm-mut2e-no-name-catlm}{CAT-LM\xspace}
\DefMacro{THead-catlm-ne2e-with-name-catlm}{CAT-LM\xspace}
\DefMacro{THead-catlm-ne2e-no-name-catlm}{CAT-LM\xspace}
\DefMacro{THead-selected-434-ne2e-with-name-ft-lora-codellama-7b}{No \stacktrace, \guardexp}
\DefMacro{THead-ne2e-with-name-ft-lora-codellama-7b}{No \stacktrace, \guardexp}
\DefMacro{THead-selected-434-ne2e-no-name-ft-lora-codellama-7b}{MUT+FILE+NE2E\xspace}
\DefMacro{THead-selected-434-ne2e-all-no-name-ft-lora-codellama-7b}{MUT+FILE+NE2E-sample5\xspace}
\DefMacro{THead-selected-434-ne2e-all-with-name-ft-lora-codellama-7b}{MUT+FILE+NE2E-sample5\xspace}
\DefMacro{THead-selected-434-nestack2e-noexlong-name-ft-lora-codellama-7b}{MUT+FILE+NE+STACK2E\xspace}
\DefMacro{THead-selected-434-nestack2e-with-name-ft-lora-codellama-7b}{MUT+FILE+NE+STACK2E\xspace}
\DefMacro{THead-selected-434-nestack2e-with-name-ft-lora-codellama-13b}{\Tool 13B\xspace}
\DefMacro{THead-conditionnestack2e-with-name-zero-shot-lora-codellama-7b}{\CodeLlama zero-shot}
\DefMacro{THead-conditionnestack2e-no-name-zero-shot-lora-codellama-7b}{\CodeLlama zero-shot}
\DefMacro{THead-selected-434-conditionnestack2e-no-name-zero-shot-lora-codellama-7b}{\CodeLlama zero-shot}
\DefMacro{THead-selected-434-conditionnestack2e-with-name-zero-shot-lora-codellama-7b}{\CodeLlama zero-shot}
\DefMacro{THead-mut2e-with-name-ft-lora-codellama-13b}{\Tool-base 13B\xspace}
\DefMacro{THead-conditionnestack2e-with-name-ft-lora-codellama-13b}{\Tool 13B\xspace}
\DefMacro{THead-conditionnestack2e-few-shot-no-name-gpt-4o}{\Tool(\GPTFour)\xspace}
\DefMacro{THead-conditionnestack2e-few-shot-with-name-gpt-4o}{\Tool--\GPTFour\xspace}
\DefMacro{THead-selected-434-nestack2e-no-name-ft-lora-codellama-13b}{MUT+FILE+NE+STACK2E(13B)\xspace}
\DefMacro{THead-selected-434-ne2e-few-shot-with-name-gpt-3.5-turbo-16k}{GPT3.5-few-shot\xspace}
\DefMacro{THead-ne2e-few-shot-with-name-gpt-4o}{\GPTFour-few-shot\xspace}
\DefMacro{THead-ne2e-few-shot-no-name-gpt-4o}{GPT4o-few-shot\xspace}
\DefMacro{THead-selected-434-ne2e-few-shot-no-name-gpt-3.5-turbo-16k}{GPT3.5-few-shot\xspace}
\DefMacro{THead-ne2e-few-shot-with-name-gpt-3.5-turbo-16k}{GPT3.5-few-shot\xspace}
\DefMacro{THead-ne2e-few-shot-no-name-gpt-3.5-turbo-16k}{GPT3.5-few-shot\xspace}
\DefMacro{THead-selected-434-conditionne2e-no-name-ft-lora-codellama-7b}{MUT+FILE+NE+CONDITION2E\xspace}
\DefMacro{THead-conditionne2e-with-name-ft-lora-codellama-7b}{No \stacktrace}
\DefMacro{THead-selected-434-conditionne2e-with-name-ft-lora-codellama-7b}{No \stacktrace}
\DefMacro{THead-selected-434-conditionnestack2e-no-name-ft-lora-codellama-7b}{\Tool\xspace}
\DefMacro{THead-selected-434-conditionnestack2e-with-name-ft-lora-codellama-7b}{\Tool\xspace}
\DefMacro{THead-selected-434-nestack2e-all-with-name-ft-lora-codellama-7b}{MUT+FILE+NE+STACK2E-sample5\xspace}
\DefMacro{THead-selected-434-nestack2e-all-no-name-ft-lora-codellama-7b}{MUT+FILE+NE+STACK2E-sample5\xspace}
\DefMacro{THead-selected-434-conditionnestack2e-all-with-name-ft-lora-codellama-7b}{\Tool-sample\xspace}
\DefMacro{THead-conditionnestack2e-all-with-name-ft-lora-codellama-7b}{\Tool-sample\xspace}
\DefMacro{THead-selected-434-conditionnestack2e-all-no-name-ft-lora-codellama-7b}{\Tool-sample\xspace}
\DefMacro{THead-conditionnestack2e-all-no-name-ft-lora-codellama-7b}{\Tool-sample\xspace}

\DefMacro{THead-mut2e-no-name-ft-lora-codellama-7b}{MUT+FILE2E\xspace}
\DefMacro{THead-ne2e-no-name-ft-lora-codellama-7b}{MUT+FILE+NE2E\xspace}
\DefMacro{THead-ne2e-all-with-name-ft-lora-codellama-7b}{MUT+FILE+NE2E-sample5\xspace}
\DefMacro{THead-ne2e-all-no-name-ft-lora-codellama-7b}{MUT+FILE+NE2E-sample5\xspace}
\DefMacro{THead-nestack2e-no-name-ft-lora-codellama-7b}{MUT+FILE+NE+STACK2E\xspace}
\DefMacro{THead-nestack2e-with-name-ft-lora-codellama-7b}{MUT+FILE+NE+STACK2E\xspace}
\DefMacro{THead-nestack2e-with-name-ft-lora-codellama-7b-real}{MUT+FILE+NE+STACK2E\xspace}
\DefMacro{THead-nestack2e-no-name-ft-lora-codellama-7b-real}{MUT+FILE+NE+STACK2E\xspace}
\DefMacro{THead-nestack2e-with-name-ft-lora-codellama-13b-real}{MUT+FILE+NE+STACK2E(13B)\xspace}
\DefMacro{THead-nestack2e-no-name-ft-lora-codellama-13b-real}{MUT+FILE+NE+STACK2E(13B)\xspace}
\DefMacro{THead-nestack2e-no-name-ft-lora-codellama-13b}{MUT+FILE+NE+STACKTRACE(13B)\xspace}
\DefMacro{THead-conditionne2e-no-name-ft-lora-codellama-7b}{MUT+FILE+NE+CONDITION2E\xspace}
\DefMacro{THead-conditionne2e-with-name-ft-lora-codellama-7b-real}{MUT+FILE+NE+CONDITION2E\xspace}
\DefMacro{THead-conditionne2e-no-name-ft-lora-codellama-7b-real}{MUT+FILE+NE+CONDITION2E\xspace}
\DefMacro{THead-conditionnestack2e-no-name-ft-lora-codellama-7b}{\Tool\xspace}
\DefMacro{THead-conditionnestack2e-with-name-ft-lora-codellama-7b}{\Tool\xspace}
\DefMacro{THead-conditionnestack2e-no-name-ft-lora-codellama-7b-real}{MUT+FILE+NE+CONDITION+STACK2E\xspace}
\DefMacro{THead-conditionnestack2e-with-name-ft-lora-codellama-7b-real}{MUT+FILE+NE+CONDITION+STACK2E\xspace}

\DefMacro{THead-mut2e-zero-shot-gpt-4-0613}{GPT4 zero-shot MUT+FILE2E\xspace}
\DefMacro{THead-mut2e-zero-shot-gpt3.5}{GPT3.5 zero-shot MUT+FILE2E\xspace}
\DefMacro{THead-mut2e-ft-axolotl-llama}{CodeLlama-7b MUT+FILE2E}
\DefMacro{THead-nestack2e-axolotl-codellama-7b}{CodeLlama-7b NESTACK2E}
\DefMacro{THead-cov2e-1-ft-axolotl-codellama-7b}{CodeLlama-7b NE2E}
\DefMacro{THead-nestack2e-ft-axolotl-llama}{CodeLlama-7b NESTACK2E\xspace}
\DefMacro{THead-nestack2e-no-stack-trace-axolotl-codellama-7b}{CodeLlama-7b (w.o. s.t.)}
\DefMacro{THead-nestack2e-with-stack-trace-axolotl-codellama-7b}{CodeLlama-7b (w. s.t.)}
\DefMacro{THead-stack2e-ft-axolotl-llama}{CodeLlama-7b STACK2E}
\DefMacro{THead-ne2e-ft-axolotl-llama}{CodeLlama-7b NE2E}
\DefMacro{THead-gpt4-nestack2e-cot-gpt-4-0613}{GPT4 COT (one-shot)\xspace}
\DefMacro{THead-gpt4-nestack2e-few-shot-gpt-4-0613}{GPT4 one-shot\xspace}

\DefMacro{THead-gt-nestack2e-few-shot-no-name-gpt-3.5-turbo-16k}{GPT3.5-gt-st-no-name\xspace}
\DefMacro{THead-gt-conditionne2e-few-shot-no-name-gpt-3.5-turbo-16k}{GPT3.5-gt-condition-no-name\xspace}
\DefMacro{THead-nestack2e-few-shot-no-name-gpt-3.5-turbo-16k}{GPT3.5-st-no-name\xspace}

\DefMacro{THead-analysis-time}{Analysis Time\xspace}
\DefMacro{THead-inference-time}{Inference Time\xspace}
\DefMacro{THead-total-time}{Total\xspace}

\DefMacro{THead-expect}{Test(expected)\xspace}
\DefMacro{THead-assertThrows}{assertThrow\xspace}
\DefMacro{THead-expectedRule}{expectedRule\xspace}
\DefMacro{THead-try{fail}catch}{tryFailCatch\xspace}

\DefMacro{THead-mut2e-stmt-otest-chatgpt-basic-top1}{ChatGPT-zero-shot\xspace}
\DefMacro{THead-mut2e-stmt-otest-SingleEvaluator-base-4197-bs10-last}{TeCo-basic\xspace}
\DefMacro{THead-mut2e-stmt-otest-chatgpt-one-shot-top1}{ChatGPT-one-shot\xspace}
\DefMacro{THead-mut2e-stmt-test-chatgpt-zero-shot-full-context-top1}{ChatGPT-zero-shot-full-context\xspace}
\DefMacro{THead-mean-stmts-toks}{len(stmts)\xspace}
\DefMacro{THead-num-estests}{\# \aETest}
\DefMacro{THead-num-projects}{\#project\xspace}
\DefMacro{THead-num-stmts}{\#stmts\xspace}

\DefMacro{THead-etest-count}{\#\EBTs\xspace}
\DefMacro{THead-test-count}{\#Tests\xspace}
\DefMacro{THead-mut-count}{\#MUTs\xspace}
\DefMacro{THead-total-test-count}{\# Tests\xspace}
\DefMacro{THead-module-count}{\#Modules\xspace}
\DefMacro{THead-etype-count}{\# Ex. Type\xspace}

\DefMacro{THead-num-tests}{\#Tests\xspace}
\DefMacro{THead-evosuite-num-tests}{\EvoSuite\xspace}
\DefMacro{THead-randoop-num-tests}{\Randoop\xspace}
\DefMacro{THead-time}{Time (s)\xspace}
\DefMacro{THead-evosuite-time}{\EvoSuite\xspace}
\DefMacro{THead-randoop-time}{\Randoop\xspace}
\DefMacro{THead-num-public-throw-stmts}{\makecell{\#Throw Statements\\in Public Methods}}
\DefMacro{THead-num-eval-throw-stmts}{\makecell{\#Throw Statements\\for Evaluation}}

\DefMacro{NumDiversitySubset}{253}
\DefMacro{TotalJavaProj}{4,767}
\DefMacro{UsedJavaProj}{}
\DefMacro{expectedPct}{65\%}
\DefMacro{tryfailPct}{24\%}
\DefMacro{realStracePct}{61\%}
\DefMacro{testSize}{842}
\DefMacro{tsEtests}{434}
\DefMacro{machineViewSize}{649}
\DefMacro{RandoopSuccProjs}{27}
\DefMacro{RandoopFailProjs}{three}
\DefMacro{EvosuiteAverageTime}{1752.6s}
\DefMacro{RandoopAverageTime}{6621.4s}
\DefMacro{exLongAverageTime}{501.3s}
\DefMacro{exLongAnalysisTime}{276.9s}
\DefMacro{exLongGenTime}{224.4s}
\DefMacro{Evosuite-all-projects-coverage}{20.95}
\DefMacro{Evosuite-subset-projects-coverage}{20.37}
\DefMacro{Randoop-all-projects-coverage}{18.95}
\DefMacro{Randoop-subset-projects-coverage}{21.87}

\newcommand{\allEvalProjects}{30}

\newcommand{\improvedPctCoverageThanCAT}{83.8\%}
\newcommand{\reducedQuant}{13.1\%}

\newcommand{\exampleMUT}{\texttt{schedule}\xspace}
\newcommand{\TSCoverage}{ThrowCov\%\xspace}
\newcommand{\runnable}{Runnable\%\xspace}
\newcommand{\compile}{Compilable\%\xspace}

\newcommand{\ToolMoreNebt}{\Tool-sample\xspace}

\newcommand{\OverGPTRun}{9.9\%}
\newcommand{\OverGPTCov}{22.8\%}
\newcommand{\OverCATRun}{83.8\%}
\newcommand{\OverCATCov}{98.0\%}

\input{tables/numbers-tools-cmp-table}
\input{tables/numbers-user-view-with-name-table}

\acmBooktitle{Companion Proceedings of the 33rd ACM Symposium on the Foundations of Software Engineering (FSE '25), June 23--27, 2025, Trondheim, Norway}
\title{\Title}

\author{Linghan Zhong}
\affiliation{
\institution{UT Austin (USA)}
\country{}
}
\email{linghanz@cs.utexas.edu}
\author{Samuel Yuan}
\affiliation{
\institution{UT Austin (USA)}
\country{}
}
\email{syuan@utexas.edu}
\author{Jiyang Zhang}
\affiliation{
\institution{UT Austin (USA)}
\country{}
}
\email{jiyang.zhang@utexas.edu}
\author{Yu Liu}
\affiliation{
\institution{UT Austin (USA)}
\country{}
}
\email{yuki.liu@utexas.edu}
\author{Pengyu Nie}
\affiliation{
\institution{UWaterloo (Canada)}
\country{}
}
\email{pynie@uwaterloo.ca}
\author{Junyi Jessy Li}
\affiliation{
\institution{UT Austin (USA)}
\country{}
}
\email{jessy@austin.utexas.edu}
\author{Milos Gligoric}
\affiliation{
\institution{UT Austin (USA)}
\country{}
}
\email{gligoric@utexas.edu}

\begin{document}

\begin{abstract}
Exceptional behavior tests (\EBTs) are crucial in software development for
verifying that code correctly handles unwanted events and throws appropriate
exceptions. However, prior research has shown that developers often prioritize
testing ``happy paths'', i.e., paths without unwanted events, over exceptional
scenarios. We present \Tool, a tool that automatically generates \EBTs to
address this gap. \Tool leverages a large language model (LLM) fine-tuned from
CodeLlama and incorporates reasoning about exception-throwing traces,
conditional expressions that guard throw statements, and non-exceptional
behavior tests that execute similar traces. Our demonstration video
illustrates how \Tool can effectively assist developers in creating
comprehensive \EBTs for their project (available at
\url{https://youtu.be/Jro8kMgplZk}).
\end{abstract}

\copyrightyear{2025}
\acmYear{2025}
\setcopyright{cc}
\setcctype{by-nd}
\acmConference[FSE Companion '25]{33rd ACM International Conference on the Foundations of Software Engineering}{June 23--28, 2025}
{Trondheim, Norway}
\acmBooktitle{33rd ACM International Conference on the Foundations of Software Engineering (FSE Companion '25), June 23--28, 2025,
Trondheim, Norway}\acmDOI{10.1145/3696630.3728608}
\acmISBN{979-8-4007-1276-0/2025/06}



\maketitle

\section{Introduction}

Many popular programming languages, including C\#, Java, and Python, support
exceptions~\cite{hejlsberg2003c, gosling2000java, vanrossum2010python}.
Exceptions are thrown during program execution if an unwanted event happens,
\eg, a method is invoked with an illegal argument value. Software developers
write \emph{exceptional behavior tests} (\EBTs) to check that their code
properly detects unwanted events and throws desired exceptions.
Prior research studies on
\EBTs~\cite{DaltonETAL20ExceptionalBehaviorTesting,
BernardoETAL11AgileTestingOfExceptionalBehavior,
GoffiETAL16AutomaticGenerationOfOraclesForExceptionalBehaviors,
MarcilioFuria21HowJavaProgrammersTestExceptionalBehavior,
LimaETAL21AssessingExceptionHandlingTesting} have shown the
importance of \EBTs and developers' desire to improve the testing of
exceptional behaviors.  However, in practice, developers tend to focus
on ``happy paths'' and have limited time to test exceptional
behaviors. This results in a lower number of \EBTs compared to \nEBTs
in most projects.

Sadly, tool support for automatically generating \EBTs is
limited. Most existing analysis-based test generation tools (\eg,
\Randoop~\cite{PachecoETAL07Randoop, RobinsonETAL11ScalingTestGen} and
\EvoSuite~\cite{FraserAndArcuri11EvoSuite}) and learning-based test
generation tools (\eg, \CAT~\cite{RaoETAL23CAT} and
\TeCo~\cite{NieETAL23TeCo}) have no special settings for targeting
\EBTs and are primarily evaluated on \nEBTs. Random test generation
tools can be guided by reinforcement learning to target exceptional
behaviors~\cite{AlmullaETAL20GeneratingExceptionTriggeringTests}, but
the generation works only on the entire codebase, and not for a
specific throw statement that a developer might select. Additionally,
tests produced by analysis-based tools often lack
readability~\cite{daka2017generating, panichella2022test,
daka2018improving}.

\begin{figure}[t]
\begin{subfigure}{\columnwidth}
\begin{lstlisting}[language=java-pretty]
class Scheduler {
  ...
  public Job schedule(String nullableName,Runnable runnable,Schedule when){(*@\label{example:mut}@*)
    ...
		Job job = prepareJob(name, runnable, when);
    ...
		return job;}
  private Job prepareJob(String name, Runnable runnable, Schedule when)
		synchronized (indexedJobsByName) {
			Job lastJob = findJob(name).orElse(null);  (*@\label{example:assignment}@*)
			if(lastJob != null && lastJob.status() != JobStatus.DONE) { (*@\label{example:condition}@*)
				throw new IllegalArgumentException("A job is already scheduled with the name:" + name); (*@\label{example:ts}@*)}
      ...
			return job;}}}
\end{lstlisting}
\vspace{-10pt}
\caption{Method under test: \CodeIn{schedule}.}
\label{fig:example-mut}
\end{subfigure}
\begin{subfigure}{\columnwidth}
\begin{lstlisting}[language=java-pretty]
@Test(expected = IllegalArgumentException.class)
public void reject_scheduling_a_job_with_same_name_but_different_runnable() {
    Scheduler scheduler = new Scheduler();
    Job j1 = scheduler.schedule("myJob", runnable1, now().plusSeconds(5));
    scheduler.schedule("myJob", runnable2, now().plusSeconds(6));}
\end{lstlisting}
\vspace{-10pt}
\caption{\EBT generated by \Tool.}
\label{fig:example-test}
\end{subfigure}
\caption{\UserView example\label{fig:example}.}
\vspace{-15pt}
\end{figure}

\begin{figure}[t]
\begin{subfigure}{\columnwidth}
\begin{lstlisting}[language=java-pretty]
public Scheduler(SchedulerConfig config) {
    if(config.getTimeProvider() == null) {
        throw new NullPointerException("The timeProvider cannot be null"); (*@\label{example:machine-ts}@*)}
    ...
}
\end{lstlisting}
\vspace{-10pt}
\caption{Method under test: \CodeIn{Scheduler}.}
\label{fig:example-machine-mut}
\end{subfigure}
\begin{subfigure}{\columnwidth}
\begin{lstlisting}[language=java-pretty]
@Test(expected = NullPointerException.class)
public void should_fail_if_timeProvider_is_null() {
    new Scheduler(SchedulerConfig.builder().maxThreads(1).timeProvider(null).build());}
\end{lstlisting}
\vspace{-10pt}
\caption{\EBT generated by \Tool.}
\label{fig:example-machine-test}
\end{subfigure}
\caption{\MachineView example\label{fig:example-machine}.}
\vspace{-15pt}
\end{figure}

We recently designed and developed \Tool~\cite{zhang2024exlong},
a framework that utilized an instruction fine-tuned large language
model (\LLM) to automatically generate \EBTs. Using
\CodeLlama~\cite{roziere2023code} as its base, \Tool is
\finetuned~\cite{selfInstruct,wei2021finetuned,sanh2021multitask} with
a novel task instruction dataset, designed specifically to embed the
reasoning about the context which includes: (a)~\stacktraces that lead
to target throw statements, (b)~\guardexps (\ie, conditional
expressions that guard those throw statements), and (c)~\nEBTs that
execute similar traces. This context is used as the input to generate
an \EBT that triggers the target throw statement.  In
figures~\ref{fig:example} and \ref{fig:example-machine}, we show
examples of \EBTs generated by \Tool.

This paper extends \Tool by introducing a new command-line interface that
simplifies the process of extracting the necessary context for \EBTs generation
and querying the fine-tuned LLM. We describe two use cases supported by \Tool:
(1)~\userView: developers select a method under test (\eg, \CodeIn{schedule} in
Figure~\ref{fig:example-mut}), a target throw statement (\eg,
line~\ref{example:ts} in Figure~\ref{fig:example-mut})
and a \testfile. \Tool then automatically generates an \EBT that executes the
target throw statement.
(2)~\machineView: developers employ \Tool to automatically generate
\EBTs for their entire codebase, covering each existing throw
statement, such as line~\ref{example:machine-ts} in \CodeIn{Scheduler}
in Figure~\ref{fig:example-machine-mut}. Additionally, to improve
\Tool's accessibility for typical users, we include an option to use a
quantized~\cite{dettmers2024int8, yao2022zero} version of the
\finetuned \LLM, which reduces the memory usage by 75\%. This
optimization enables \Tool to operate on machines with limited
computational resources.

Our experiments demonstrate \Tool's effectiveness in both supported use cases.
For the \userView, we compare our tool against a state-of-the-art test
generation model (\CAT~\cite{RaoETAL23CAT}) and a leading foundation \LLM
(\GPTThree~\cite{gpt3.5-turbo}). Results show that \Tool generates
\improvedPctCoverageThanCAT{} more executable \EBTs than \CAT and \OverGPTRun{}
more than \GPTThree.
After quantization, \Tool can run on a local machine with a single GPU, with a
relative small performance reduction resulting in the generation of
\reducedQuant{} fewer executable \EBTs. For the machine-oriented use case, we
compare our tool against two popular analysis-based test generation tools:
\Randoop~\cite{PachecoETAL07Randoop, RobinsonETAL11ScalingTestGen} and
\EvoSuite~\cite{FraserAndArcuri11EvoSuite}. While these tools complement each
other (\ie, each tool can generate \EBTs for some target throw statements that
others cannot), our findings indicate that \Tool outperforms both \Randoop and
\EvoSuite.
\Tool is available on GitHub at
\url{https://github.com/EngineeringSoftware/exLong}.

\section{Technique and Implementation}
\label{sec:tech}

\begin{figure*}[t]
\centering \includegraphics[width=0.75\textwidth]{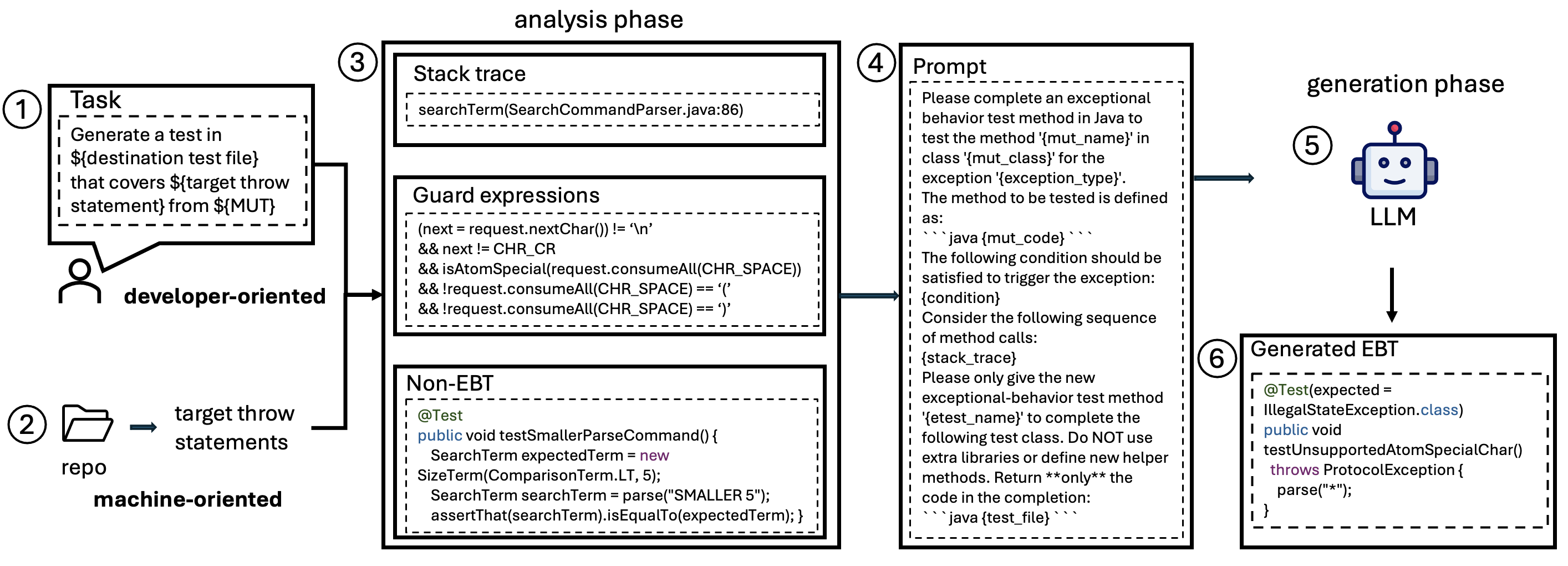}
\vspace{-10pt}
\caption{Overview of \Tool.\label{fig:overview}}
\vspace{-10pt}
\end{figure*}

\begin{figure}[t]
\begin{subfigure}{\columnwidth}
\begin{lstlisting}[language=java-pretty]
schedule(Scheduler.java:186)
    ...
    Job job = prepareJob(name, runnable, when);
    ...
prepareJob(Scheduler.java:340)
    ...
    throw new IllegalArgumentException("A job is already scheduled with the name:" + name);
    ...
\end{lstlisting}
\vspace{-1pt}
\caption{Stack trace from \MUT to \ts.}
\label{fig:example-stack}
\end{subfigure}
\begin{subfigure}{\columnwidth}
\begin{lstlisting}[language=java-pretty]
findJob(nullableName == null ? runnable.toString() : nullableName).orElse(null) != null && findJob(nullableName == null ? runnable.toString() : nullableName).orElse(null).status() != JobStatus.DONE
\end{lstlisting}
\vspace{-1pt}
\caption{Guard expression.}
\label{fig:example-condition}
\end{subfigure}
\begin{subfigure}{\columnwidth}
\begin{lstlisting}[language=java-pretty]
@Test
public void should_run_a_single_job() throws InterruptedException {
    Scheduler scheduler = new Scheduler();
    SingleJob singleJob = new SingleJob();
    scheduler.schedule("test", singleJob, Schedules.executeOnce(Schedules.fixedDelaySchedule(Duration.ofMillis(1))));
    waitOn(singleJob, () -> singleJob.countExecuted.get() > 0, 10000);
    scheduler.gracefullyShutdown();
    assertThat(singleJob.countExecuted.get()).isEqualTo(1);}
\end{lstlisting}
\vspace{-1pt}
\caption{\nEBT.}
\vspace{-7pt}
\label{fig:example-ne}
\end{subfigure}
\caption{Context for \Tool.\label{fig:analysis}}
\end{figure}

Figure~\ref{fig:overview}~\citep{zhang2024exlong} illustrates the workflow of
\Tool. Given a method under test (\MUT), a \ts, and a \testfile , \Tool collects
\stacktrace, \guardexp, and relevant \nEBTs using both static and dynamic
program analyses (\circled{3}). These components are then used to construct a
prompt which encompasses both the task inputs and the relevant context
(\circled{4}). During training, a foundation \LLM is \finetuned to generate the
\EBT conditioned on the input $\aPrompt$. During inference, \Tool first prepares
the necessary context to construct the prompt then the fine-tuned \LLM generates
\EBTs given the prompt. We detail the design and implementation in the rest of
this section.

\subsection{\UserView}

\label{sec:tech-userview}

\noindent \MyPara{Preparation} In this phase, \Tool collects a set of
\stacktraces from the execution of existing \nEBTs, that can reach methods that
contain \tss in the \repo.%
Using the example in Figure~\ref{fig:example}, \Tool first identifies and
instruments the throw statement in the method \Code{prepareJob} to log the
current \stacktrace upon the invocation of \CodeIn{prepareJob}. Then \Tool
executes the existing \nEBTs to log the \stacktraces and record the mapping
between the non-EBTs and their invoked methods.
Note that a developer only need to run this phase once for the repository they
are working on.

\MyPara{Analysis} \Tool constructs a prompt from the developer-provided context
and the information collected in the preparation phase.
Taking Figure~\ref{fig:example} as an
example, \Tool first searches the collected \stacktraces for one that begins
with \exampleMUT and ends in \CodeIn{prepareJob}. An example of the resulting
\stacktrace consisting of the \exampleMUT and \CodeIn{prepareJob} methods is
shown in Figure~\ref{fig:example-stack}. While \stacktrace provides the sequence
of method invocations
that lead to the \ts, knowing only the names of the methods is insufficient for
generating \EBTs. \Tool then constructs a
\guardexp to further aid the \LLM's reasoning about system configurations
that would lead to exceptional behaviors. A \guardexp is a logical formula representing the constraints necessary to
reach the \ts. An example of \guardexp is shown in
Figure~\ref{fig:example-condition}. Specifically, \Tool collects guard-related
AST nodes along the \stacktrace, including conditional expressions
(line~\ref{example:condition} in Figure~\ref{fig:example}) and assignments
(line~\ref{example:assignment} in Figure~\ref{fig:example}). It then propagates
symbolic variables, performing substitutions where
necessary. %
The resulting formula is a conjunction of expressions guarding the target throw
statement. Finally, \Tool identifies relevant \nEBTs from the same repository to encourage the \LLM to reason about the procedures to set up the
object under test and to promote consistency between the newly generated code and
existing code in terms of format and coding conventions. The \nEBT in
figure~\ref{fig:example-ne} is identified as relevant since it invokes the
target \MUT \exampleMUT.
To enhance the quality of the generated \EBTs, \Tool can optionally create
multiple prompts by including different relevant \nEBTs and then select the best
\EBT based on its ability to compile, execute, and cover the target throw
statement.

\subsection{\MachineView}

\MyPara{Preparation}
\Tool parses the repository to identify all \tss within public methods
(line~\ref{example:machine-ts} in Figure~\ref{fig:example-machine}). Similar to
\userView, it executes the existing \nEBTs to extract the coverage data.
This is used to determine both the relevant \nEBTs and the destination test
file.

\MyPara{Analysis} As shown in Figure~\ref{fig:example-machine-mut}, for each
\ts, the \MUT is defined as the method containing the \ts (\CodeIn{Scheduler}).
In this case, the \stacktrace only includes the \MUT.
The \guardexp and relevant \nEBTs are extracted using the same approach as
\userView. The \testfile is selected using two heuristics similar to
prior works~\cite{RaoETAL23CAT}: (1) file name matching where given a code file named
\texttt{Scheduler.java}, \Tool searches for test file named
\texttt{TestScheduler.java} or \texttt{SchedulerTest.java}, and (2) test coverage
analysis in which if name matching fails, \Tool searches for the test class covering the
\MUT or the class of the \MUT.
Finally, \Tool constructs the prompt with all the available context. \Tool can
optionally create multiple prompts from different \nEBTs, generating and
evaluating multiple \EBTs then select the best one based on runtime evaluation.

\section{Tool Installation}

\Tool generates \EBTs for \Java projects built using Maven. We require
Maven 3.8.3+ and Java 8+. For quantized \LLM inference,
\Tool leverages \ollama~\cite{yang2025ollama}, which can be installed
following the instructions from \ollama's official GitHub repository.

To get started with \Tool, begin by cloning the repository:
\begin{lstlisting}[language=bash,basicstyle=\scriptsize\ttfamily]
$ git clone https://github.com/EngineeringSoftware/exLong.git
\end{lstlisting}
\Tool is implemented in \python and requires version 3.10 or
higher. For a smooth installation process, we recommend using Conda
\cite{Conda} to manage dependencies. Users can execute our provided script to
set up \Tool and its required components:

\begin{lstlisting}[language=bash,basicstyle=\scriptsize\ttfamily]
$ ./scripts/prepare_conda_env.sh
\end{lstlisting}
We also offer Docker-based installation options. The Docker
image can be built and run with:
\begin{lstlisting}[language=bash,basicstyle=\scriptsize\ttfamily]
$ docker build -t exlong .
$ docker exec -it exlong /bin/bash
\end{lstlisting}
Furthermore, for integration with the \ollama Docker image, the users can use our
Docker Compose setup:
\begin{lstlisting}[language=bash,basicstyle=\scriptsize\ttfamily]
$ docker compose up -d
$ docker exec -it exlong-tool-1 /bin/bash
\end{lstlisting}

\section{Tool Usage}

In this section, we introduce how to use \Tool for \userView and
\machineView.

\subsection{\UserView}

For the \userView, where \Tool generates an \EBT for a user-specified
target throw statement, our tool's CLI requires the following parameters:
the local path or remote link to the git repository, the path to the
file containing the \MUT, the line number of the beginning of \MUT's
definition, the path to the file containing the \ts, the line number
of the \ts, and the path to the \testfile.

Additionally, \Tool's CLI accepts the following optional parameters: a
commit SHA (default: latest commit on the main branch), name of the
test method to be written by \Tool (default: none), whether \Tool should used quantized \LLM (default: true), whether \Tool
should sample multiple candidate \EBTs and select the best test based
on runtime evaluation (default: false), and the output file path for
the generated \EBT (default: \CodeIn{./output.java}).

An example command to invoke \userView of \Tool is as follows:

\begin{lstlisting}[language=bash,basicstyle=\scriptsize\ttfamily]
$ python -m etestgen.cli user_view \
--repo_path=./Wisp \
--mut_file_path=Scheduler.java \
--mut_line=180 \
--quant=true \
--throw_file_path=Scheduler.java \
--throw_line=340 \
--test_context_path=SchedulerTest.java
--sha="ce1d9f3cb1944115ad98b4428ea24b24ab3faf56" \
--test_name=testSchedulerError \
--pick_best=True \
--output_file=./ExlongTest.java
\end{lstlisting}

\subsection{\MachineView}

In the \machineView, where \Tool generates \EBTs for the entire
codebase. The only required parameter for \Tool's CLI is the path or link to
the git repository. The CLI also accepts commit SHA, option to sample
multiple \EBTs, option to use quantized \LLM, time budget
for \Tool to finish, and path to output file
as optional parameters.

An example command to invoke \userView of \Tool is as follows:
\begin{lstlisting}[language=bash,basicstyle=\scriptsize\ttfamily]
$ python -m etestgen.cli machine_view  \
--repo_link= \
"https://github.com/Coreoz/Wisp.git" \
--sha="ce1d9f3cb1944115ad98b4428ea24b24ab3faf56" \
--timeout=1000
\end{lstlisting}

\section{Evaluation}

\begin{table}[t]
\begin{small}
\begin{center}
\caption{\UseMacro{TCap-models-user-view-with-name}\label{tab:user-view-with-name}\vspace{-5pt}}
\begin{tabular}{l | c c c }
\toprule
\textbf{\UseMacro{THead-models}}
& \textbf{\UseMacro{THead-compilable-max}}
& \textbf{\UseMacro{THead-runnable-overall-max}}
& \textbf{\UseMacro{THead-coverage-max}}
\\
\midrule
\UseMacro{THead-ne2e-few-shot-with-name-gpt-3.5-turbo-16k}
& \UseMacro{res-user-view-with-name-ne2e-few-shot-with-name-gpt-3.5-turbo-16k-compilable-max}
& \UseMacro{res-user-view-with-name-ne2e-few-shot-with-name-gpt-3.5-turbo-16k-runnable-overall-max}
& \UseMacro{res-user-view-with-name-ne2e-few-shot-with-name-gpt-3.5-turbo-16k-coverage-max}
\\
\UseMacro{THead-catlm-ne2e-with-name-catlm}
& \UseMacro{res-user-view-with-name-catlm-ne2e-with-name-catlm-compilable-max}
& \UseMacro{res-user-view-with-name-catlm-ne2e-with-name-catlm-runnable-overall-max}
& \UseMacro{res-user-view-with-name-catlm-ne2e-with-name-catlm-coverage-max}
\\
\UseMacro{THead-conditionnestack2e-with-name-ft-lora-codellama-7b}
& \textbf{\UseMacro{res-user-view-with-name-conditionnestack2e-with-name-ft-lora-codellama-7b-compilable-max}}
& \textbf{\UseMacro{res-user-view-with-name-conditionnestack2e-with-name-ft-lora-codellama-7b-runnable-overall-max}}
& \textbf{\UseMacro{res-user-view-with-name-conditionnestack2e-with-name-ft-lora-codellama-7b-coverage-max}}
\\
\bottomrule
\end{tabular}
\end{center}
\end{small}
\vspace{-10pt}
\end{table}

Following prior work~\cite{NieETAL23TeCo}, we collect our dataset from Java
projects in CodeSearchNet~\cite{CodeSearchNet}, which are available on GitHub.
We evaluate \Tool's performance with full
precision \LLM under both \userView and \machineView. For \userView, we benchmark \Tool on a subset of
\UseMacro{tsEtests} examples from which we are able to extract stack traces. For
\machineView, we evaluate \Tool on \UseMacro{machineViewSize} examples,
filtering out data for which our heuristic failed to locate the corresponding
\testfile.

We evaluate \EBTs generated by \Tool using the percentage of generated \EBTs
that can be compiled (\compile),
can be executed (Runnable\%), and those that are semantically valid and are
targeting the throw statement specified by developers (\TSCoverage). We compare
\Tool against a widely used foundation model, \GPTThree, and a specialized
test-generating LLM, \CAT.
Our results are shown in Table~\ref{tab:user-view-with-name}. We observe that
\Tool outperforms all the baselines on all metrics.
\Tool achieves higher performance for both generating executable \EBTs
(Runnable\%) and \EBTs that cover the \tss (\TSCoverage). Specifically, \Tool
outperforms \GPTThree by \OverGPTRun{} and \OverGPTCov{} on \runnable and
\TSCoverage, respectively. Similarly, \Tool outperforms CAT-LM by \OverCATRun{}
and \OverCATCov{} on \runnable and \TSCoverage, respectively.

\begin{figure}[t!]
\centering
\includegraphics[width=0.25\textwidth]{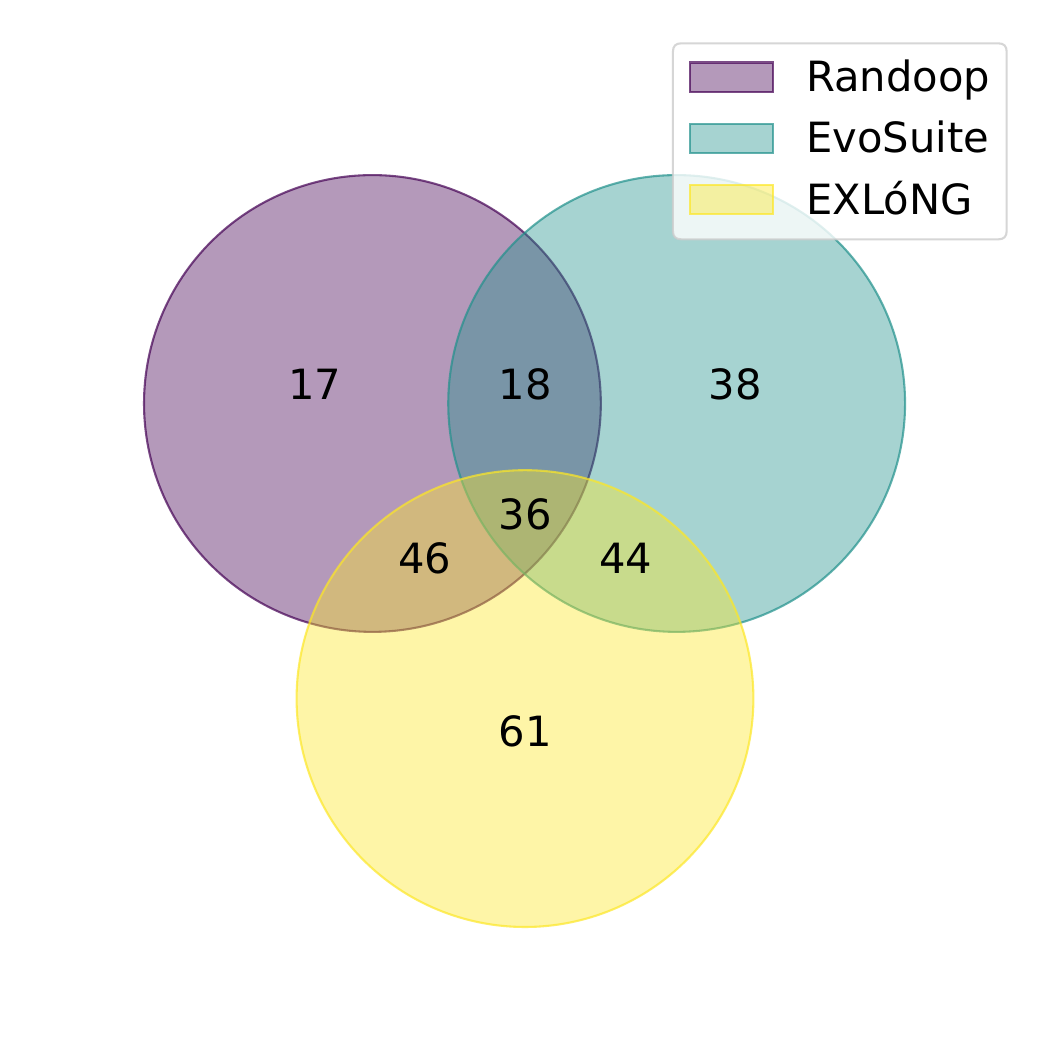}
\vspace{-20pt}
\caption{Venn diagram of \tss coverage by \Tool, \Randoop, and
\EvoSuite on all \allEvalProjects{} projects. \label{fig:Venn}}
\vspace{-5pt}
\end{figure}

For \machineView, we evaluate the tool's ability to cover throw statements
within a given repository with \TSCoverage, which measures the percentage of
\tss covered by the generated \EBTs. We benchmark \Tool against two widely-used
analysis-based test generation tools:
\Randoop~\cite{PachecoETAL07Randoop,RobinsonETAL11ScalingTestGen} and
\EvoSuite~\cite{FraserAndArcuri11EvoSuite}. Our results, illustrated in
Figure~\ref{fig:Venn}, indicates that \Tool covers the most \tss. For more details of our evaluation, refer to the full
paper~\cite{zhang2024exlong}.

\section{Related Work}
\label{sec:related:work}

Recent studies have leveraged transformer models for test
generation~\cite{wang2024software, el2024using, SchaferETAL23TestPilot,
TufanoETAL20TestGeneration, RaoETAL23CAT, NieETAL23TeCo, Nie23Thesis,
LemieuxETAL23CodaMosa, yuan2023no}. Some approaches use conditions to guide
the generation process~\cite{bouzenia2023say, blasi2022call, ryan2024code},
while others utilize existing test cases as context~\cite{RaoETAL23CAT,
TufanoETAL20TestGeneration, NieETAL23TeCo, el2024using}. Our work uniquely
combines non-exceptional tests with stack traces and \guardexp to guide
exceptional test generation.

Non-LLM test generation approaches include
random-based~\cite{PachecoETAL07Randoop, RobinsonETAL11ScalingTestGen},
search-based~\cite{FraserAndArcuri11EvoSuite, harman2009theoretical,
LiuISSTA23EXLI, LiuFSE24EXLI}, and constraint-based~\cite{ernst2007daikon,
HolmesETAL20TestGeneration, godefroid2012test} strategies. While tools like
\Randoop and \EvoSuite can generate tests for exceptional behaviors, they
neither guarantee coverage of specific exceptional paths nor consistently
produce readable test cases due to their random nature.

\section{Conclusion}
\vspace{-3pt}

We presented \Tool, a novel command-line tool that leverages large language
models to generate exceptional behavior tests (\EBTs). \Tool offers two
practical use cases: \userView, which generates an \EBT for a specific method
and \ts, and \machineView, which automatically creates tests for all \tss in a
repository.
To make \Tool accessible to general users, we provide an option which uses a
quantized \finetuned \LLM to reduce the computational cost of running inference.
We believe that \Tool targets an important task in software testing and
demonstrated strong performance. By simplifying the process of generating tests,
\Tool enables developers to more easily create comprehensive test suites that
cover exceptional behaviors.

\section*{Acknowledgments}

We thank Nader Al Awar, Jayanth Srinivasa, Aditya Thimmaiah, Zijian Yi
and Zhiqiang Zang for their feedback.  We acknowledge the Texas
Advanced Computing Center
at The University of Texas at Austin and Digital Research Alliance of
Canada
for providing HPC resources that have contributed to the research
results reported within this paper.  This work is partially supported
by the US National Science Foundation under Grant Nos.~CCF-2107291,
CCF-2217696, CCF-2313027, CCF-2403036; as well as AST-2421782 and
Simons Foundation MPS-AI-00010515 (NSF-Simons AI Institute for Cosmic
Origins -- CosmicAI).
This work was in part supported by Cisco Research.  Any opinions,
findings and conclusions, or recommendations expressed in this
material are those of the authors and do not necessarily reflect the
views of Cisco Research.

\balance
\bibliography{bib}

\end{document}

%% file: tables/numbers-tools-cmp-table.tex
\DefMacro{res-tools-cmp-conditionnestack2e-all-no-name-ft-lora-codellama-7b-eval-rq2-compilable-avg}{66.22}
\DefMacro{res-tools-cmp-conditionnestack2e-all-no-name-ft-lora-codellama-7b-eval-rq2-compilable-max}{80.43}
\DefMacro{res-tools-cmp-conditionnestack2e-all-no-name-ft-lora-codellama-7b-eval-rq2-compilable-min}{51.62}
\DefMacro{res-tools-cmp-conditionnestack2e-all-no-name-ft-lora-codellama-7b-eval-rq2-coverage-avg}{25.28}
\DefMacro{res-tools-cmp-conditionnestack2e-all-no-name-ft-lora-codellama-7b-eval-rq2-coverage-max}{28.81}
\DefMacro{res-tools-cmp-conditionnestack2e-all-no-name-ft-lora-codellama-7b-eval-rq2-coverage-min}{19.88}
\DefMacro{res-tools-cmp-conditionnestack2e-all-no-name-ft-lora-codellama-7b-eval-rq2-match-avg}{99.43}
\DefMacro{res-tools-cmp-conditionnestack2e-all-no-name-ft-lora-codellama-7b-eval-rq2-match-max}{99.43}
\DefMacro{res-tools-cmp-conditionnestack2e-all-no-name-ft-lora-codellama-7b-eval-rq2-match-min}{99.43}
\DefMacro{res-tools-cmp-conditionnestack2e-all-no-name-ft-lora-codellama-7b-eval-rq2-runnable-avg}{56.50}
\DefMacro{res-tools-cmp-conditionnestack2e-all-no-name-ft-lora-codellama-7b-eval-rq2-runnable-max}{64.16}
\DefMacro{res-tools-cmp-conditionnestack2e-all-no-name-ft-lora-codellama-7b-eval-rq2-runnable-min}{47.78}
\DefMacro{res-tools-cmp-conditionnestack2e-all-no-name-ft-lora-codellama-7b-eval-rq2-runnable-overall}{51.31}
\DefMacro{res-tools-cmp-conditionnestack2e-all-no-name-ft-lora-codellama-7b-eval-rq2-runnable-overall-avg}{40.20}
\DefMacro{res-tools-cmp-conditionnestack2e-all-no-name-ft-lora-codellama-7b-eval-rq2-runnable-overall-max}{51.31}
\DefMacro{res-tools-cmp-conditionnestack2e-all-no-name-ft-lora-codellama-7b-eval-rq2-runnable-overall-min}{29.12}
\DefMacro{res-tools-cmp-conditionnestack2e-all-no-name-ft-lora-codellama-7b-eval-rq2-timeout-avg}{2.75}
\DefMacro{res-tools-cmp-conditionnestack2e-all-no-name-ft-lora-codellama-7b-eval-rq2-timeout-max}{3.28}
\DefMacro{res-tools-cmp-conditionnestack2e-all-no-name-ft-lora-codellama-7b-eval-rq2-timeout-min}{2.50}
\DefMacro{res-tools-cmp-conditionnestack2e-all-no-name-ft-lora-codellama-7b-eval-rq2-intersect-compilable-avg}{64.97}
\DefMacro{res-tools-cmp-conditionnestack2e-all-no-name-ft-lora-codellama-7b-eval-rq2-intersect-compilable-max}{80.37}
\DefMacro{res-tools-cmp-conditionnestack2e-all-no-name-ft-lora-codellama-7b-eval-rq2-intersect-compilable-min}{50.28}
\DefMacro{res-tools-cmp-conditionnestack2e-all-no-name-ft-lora-codellama-7b-eval-rq2-intersect-coverage-avg}{25.87}
\DefMacro{res-tools-cmp-conditionnestack2e-all-no-name-ft-lora-codellama-7b-eval-rq2-intersect-coverage-max}{29.72}
\DefMacro{res-tools-cmp-conditionnestack2e-all-no-name-ft-lora-codellama-7b-eval-rq2-intersect-coverage-min}{20.19}
\DefMacro{res-tools-cmp-conditionnestack2e-all-no-name-ft-lora-codellama-7b-eval-rq2-intersect-match-avg}{100.00}
\DefMacro{res-tools-cmp-conditionnestack2e-all-no-name-ft-lora-codellama-7b-eval-rq2-intersect-match-max}{100.00}
\DefMacro{res-tools-cmp-conditionnestack2e-all-no-name-ft-lora-codellama-7b-eval-rq2-intersect-match-min}{100.00}
\DefMacro{res-tools-cmp-conditionnestack2e-all-no-name-ft-lora-codellama-7b-eval-rq2-intersect-runnable-avg}{54.52}
\DefMacro{res-tools-cmp-conditionnestack2e-all-no-name-ft-lora-codellama-7b-eval-rq2-intersect-runnable-max}{61.86}
\DefMacro{res-tools-cmp-conditionnestack2e-all-no-name-ft-lora-codellama-7b-eval-rq2-intersect-runnable-min}{46.05}
\DefMacro{res-tools-cmp-conditionnestack2e-all-no-name-ft-lora-codellama-7b-eval-rq2-intersect-runnable-overall}{49.72}
\DefMacro{res-tools-cmp-conditionnestack2e-all-no-name-ft-lora-codellama-7b-eval-rq2-intersect-runnable-overall-avg}{38.42}
\DefMacro{res-tools-cmp-conditionnestack2e-all-no-name-ft-lora-codellama-7b-eval-rq2-intersect-runnable-overall-max}{49.72}
\DefMacro{res-tools-cmp-conditionnestack2e-all-no-name-ft-lora-codellama-7b-eval-rq2-intersect-runnable-overall-min}{27.85}
\DefMacro{res-tools-cmp-conditionnestack2e-all-no-name-ft-lora-codellama-7b-eval-rq2-intersect-timeout-avg}{3.32}
\DefMacro{res-tools-cmp-conditionnestack2e-all-no-name-ft-lora-codellama-7b-eval-rq2-intersect-timeout-max}{3.95}
\DefMacro{res-tools-cmp-conditionnestack2e-all-no-name-ft-lora-codellama-7b-eval-rq2-intersect-timeout-min}{3.02}

%% file: tables/numbers-user-view-with-name-table.tex
\DefMacro{res-user-view-with-name-ne2e-few-shot-with-name-gpt-3.5-turbo-16k-bleu}{56.61}
\DefMacro{res-user-view-with-name-ne2e-few-shot-with-name-gpt-3.5-turbo-16k-bleu-avg}{56.61}
\DefMacro{res-user-view-with-name-ne2e-few-shot-with-name-gpt-3.5-turbo-16k-bleu-max}{56.61}
\DefMacro{res-user-view-with-name-ne2e-few-shot-with-name-gpt-3.5-turbo-16k-bleu-min}{56.61}
\DefMacro{res-user-view-with-name-ne2e-few-shot-with-name-gpt-3.5-turbo-16k-code-bleu}{64.28}
\DefMacro{res-user-view-with-name-ne2e-few-shot-with-name-gpt-3.5-turbo-16k-code-bleu-avg}{64.28}
\DefMacro{res-user-view-with-name-ne2e-few-shot-with-name-gpt-3.5-turbo-16k-code-bleu-max}{64.28}
\DefMacro{res-user-view-with-name-ne2e-few-shot-with-name-gpt-3.5-turbo-16k-code-bleu-min}{64.28}
\DefMacro{res-user-view-with-name-ne2e-few-shot-with-name-gpt-3.5-turbo-16k-edit-sim}{82.30}
\DefMacro{res-user-view-with-name-ne2e-few-shot-with-name-gpt-3.5-turbo-16k-edit-sim-avg}{82.30}
\DefMacro{res-user-view-with-name-ne2e-few-shot-with-name-gpt-3.5-turbo-16k-edit-sim-max}{82.30}
\DefMacro{res-user-view-with-name-ne2e-few-shot-with-name-gpt-3.5-turbo-16k-edit-sim-min}{82.30}
\DefMacro{res-user-view-with-name-ne2e-few-shot-with-name-gpt-3.5-turbo-16k-rouge-f}{76.55}
\DefMacro{res-user-view-with-name-ne2e-few-shot-with-name-gpt-3.5-turbo-16k-rouge-f-avg}{76.55}
\DefMacro{res-user-view-with-name-ne2e-few-shot-with-name-gpt-3.5-turbo-16k-rouge-f-max}{76.55}
\DefMacro{res-user-view-with-name-ne2e-few-shot-with-name-gpt-3.5-turbo-16k-rouge-f-min}{76.55}
\DefMacro{res-user-view-with-name-ne2e-few-shot-with-name-gpt-3.5-turbo-16k-rouge-p}{77.39}
\DefMacro{res-user-view-with-name-ne2e-few-shot-with-name-gpt-3.5-turbo-16k-rouge-p-avg}{77.39}
\DefMacro{res-user-view-with-name-ne2e-few-shot-with-name-gpt-3.5-turbo-16k-rouge-p-max}{77.39}
\DefMacro{res-user-view-with-name-ne2e-few-shot-with-name-gpt-3.5-turbo-16k-rouge-p-min}{77.39}
\DefMacro{res-user-view-with-name-ne2e-few-shot-with-name-gpt-3.5-turbo-16k-rouge-r}{78.15}
\DefMacro{res-user-view-with-name-ne2e-few-shot-with-name-gpt-3.5-turbo-16k-rouge-r-avg}{78.15}
\DefMacro{res-user-view-with-name-ne2e-few-shot-with-name-gpt-3.5-turbo-16k-rouge-r-max}{78.15}
\DefMacro{res-user-view-with-name-ne2e-few-shot-with-name-gpt-3.5-turbo-16k-rouge-r-min}{78.15}
\DefMacro{res-user-view-with-name-ne2e-few-shot-with-name-gpt-3.5-turbo-16k-xmatch}{14.98}
\DefMacro{res-user-view-with-name-ne2e-few-shot-with-name-gpt-3.5-turbo-16k-xmatch-avg}{14.98}
\DefMacro{res-user-view-with-name-ne2e-few-shot-with-name-gpt-3.5-turbo-16k-xmatch-max}{14.98}
\DefMacro{res-user-view-with-name-ne2e-few-shot-with-name-gpt-3.5-turbo-16k-xmatch-min}{14.98}
\DefMacro{res-user-view-with-name-ne2e-few-shot-with-name-gpt-3.5-turbo-16k-xmatch-top1}{14.98}
\DefMacro{res-user-view-with-name-ne2e-few-shot-with-name-gpt-3.5-turbo-16k-compilable-avg}{75.12}
\DefMacro{res-user-view-with-name-ne2e-few-shot-with-name-gpt-3.5-turbo-16k-compilable-max}{75.12}
\DefMacro{res-user-view-with-name-ne2e-few-shot-with-name-gpt-3.5-turbo-16k-compilable-min}{75.12}
\DefMacro{res-user-view-with-name-ne2e-few-shot-with-name-gpt-3.5-turbo-16k-coverage-avg}{48.39}
\DefMacro{res-user-view-with-name-ne2e-few-shot-with-name-gpt-3.5-turbo-16k-coverage-max}{48.39}
\DefMacro{res-user-view-with-name-ne2e-few-shot-with-name-gpt-3.5-turbo-16k-coverage-min}{48.39}
\DefMacro{res-user-view-with-name-ne2e-few-shot-with-name-gpt-3.5-turbo-16k-match-avg}{100.00}
\DefMacro{res-user-view-with-name-ne2e-few-shot-with-name-gpt-3.5-turbo-16k-match-max}{100.00}
\DefMacro{res-user-view-with-name-ne2e-few-shot-with-name-gpt-3.5-turbo-16k-match-min}{100.00}
\DefMacro{res-user-view-with-name-ne2e-few-shot-with-name-gpt-3.5-turbo-16k-runnable-avg}{81.60}
\DefMacro{res-user-view-with-name-ne2e-few-shot-with-name-gpt-3.5-turbo-16k-runnable-max}{81.60}
\DefMacro{res-user-view-with-name-ne2e-few-shot-with-name-gpt-3.5-turbo-16k-runnable-min}{81.60}
\DefMacro{res-user-view-with-name-ne2e-few-shot-with-name-gpt-3.5-turbo-16k-runnable-overall}{61.29}
\DefMacro{res-user-view-with-name-ne2e-few-shot-with-name-gpt-3.5-turbo-16k-runnable-overall-avg}{61.29}
\DefMacro{res-user-view-with-name-ne2e-few-shot-with-name-gpt-3.5-turbo-16k-runnable-overall-max}{61.29}
\DefMacro{res-user-view-with-name-ne2e-few-shot-with-name-gpt-3.5-turbo-16k-runnable-overall-min}{61.29}
\DefMacro{res-user-view-with-name-ne2e-few-shot-with-name-gpt-3.5-turbo-16k-timeout-avg}{0.00}
\DefMacro{res-user-view-with-name-ne2e-few-shot-with-name-gpt-3.5-turbo-16k-timeout-max}{0.00}
\DefMacro{res-user-view-with-name-ne2e-few-shot-with-name-gpt-3.5-turbo-16k-timeout-min}{0.00}
\DefMacro{res-user-view-with-name-catlm-ne2e-with-name-catlm-bleu}{53.49}
\DefMacro{res-user-view-with-name-catlm-ne2e-with-name-catlm-bleu-avg}{53.49}
\DefMacro{res-user-view-with-name-catlm-ne2e-with-name-catlm-bleu-max}{53.49}
\DefMacro{res-user-view-with-name-catlm-ne2e-with-name-catlm-bleu-min}{53.49}
\DefMacro{res-user-view-with-name-catlm-ne2e-with-name-catlm-code-bleu}{59.79}
\DefMacro{res-user-view-with-name-catlm-ne2e-with-name-catlm-code-bleu-avg}{59.79}
\DefMacro{res-user-view-with-name-catlm-ne2e-with-name-catlm-code-bleu-max}{59.79}
\DefMacro{res-user-view-with-name-catlm-ne2e-with-name-catlm-code-bleu-min}{59.79}
\DefMacro{res-user-view-with-name-catlm-ne2e-with-name-catlm-edit-sim}{78.91}
\DefMacro{res-user-view-with-name-catlm-ne2e-with-name-catlm-edit-sim-avg}{78.91}
\DefMacro{res-user-view-with-name-catlm-ne2e-with-name-catlm-edit-sim-max}{78.91}
\DefMacro{res-user-view-with-name-catlm-ne2e-with-name-catlm-edit-sim-min}{78.91}
\DefMacro{res-user-view-with-name-catlm-ne2e-with-name-catlm-rouge-f}{74.63}
\DefMacro{res-user-view-with-name-catlm-ne2e-with-name-catlm-rouge-f-avg}{74.63}
\DefMacro{res-user-view-with-name-catlm-ne2e-with-name-catlm-rouge-f-max}{74.63}
\DefMacro{res-user-view-with-name-catlm-ne2e-with-name-catlm-rouge-f-min}{74.63}
\DefMacro{res-user-view-with-name-catlm-ne2e-with-name-catlm-rouge-p}{77.32}
\DefMacro{res-user-view-with-name-catlm-ne2e-with-name-catlm-rouge-p-avg}{77.32}
\DefMacro{res-user-view-with-name-catlm-ne2e-with-name-catlm-rouge-p-max}{77.32}
\DefMacro{res-user-view-with-name-catlm-ne2e-with-name-catlm-rouge-p-min}{77.32}
\DefMacro{res-user-view-with-name-catlm-ne2e-with-name-catlm-rouge-r}{75.08}
\DefMacro{res-user-view-with-name-catlm-ne2e-with-name-catlm-rouge-r-avg}{75.08}
\DefMacro{res-user-view-with-name-catlm-ne2e-with-name-catlm-rouge-r-max}{75.08}
\DefMacro{res-user-view-with-name-catlm-ne2e-with-name-catlm-rouge-r-min}{75.08}
\DefMacro{res-user-view-with-name-catlm-ne2e-with-name-catlm-xmatch}{9.83}
\DefMacro{res-user-view-with-name-catlm-ne2e-with-name-catlm-xmatch-avg}{9.83}
\DefMacro{res-user-view-with-name-catlm-ne2e-with-name-catlm-xmatch-max}{9.83}
\DefMacro{res-user-view-with-name-catlm-ne2e-with-name-catlm-xmatch-min}{9.83}
\DefMacro{res-user-view-with-name-catlm-ne2e-with-name-catlm-xmatch-top1}{9.83}
\DefMacro{res-user-view-with-name-catlm-ne2e-with-name-catlm-compilable-avg}{71.83}
\DefMacro{res-user-view-with-name-catlm-ne2e-with-name-catlm-compilable-max}{71.83}
\DefMacro{res-user-view-with-name-catlm-ne2e-with-name-catlm-compilable-min}{71.83}
\DefMacro{res-user-view-with-name-catlm-ne2e-with-name-catlm-coverage-avg}{30.03}
\DefMacro{res-user-view-with-name-catlm-ne2e-with-name-catlm-coverage-max}{30.03}
\DefMacro{res-user-view-with-name-catlm-ne2e-with-name-catlm-coverage-min}{30.03}
\DefMacro{res-user-view-with-name-catlm-ne2e-with-name-catlm-match-avg}{100.00}
\DefMacro{res-user-view-with-name-catlm-ne2e-with-name-catlm-match-max}{100.00}
\DefMacro{res-user-view-with-name-catlm-ne2e-with-name-catlm-match-min}{100.00}
\DefMacro{res-user-view-with-name-catlm-ne2e-with-name-catlm-runnable-avg}{59.26}
\DefMacro{res-user-view-with-name-catlm-ne2e-with-name-catlm-runnable-max}{59.26}
\DefMacro{res-user-view-with-name-catlm-ne2e-with-name-catlm-runnable-min}{59.26}
\DefMacro{res-user-view-with-name-catlm-ne2e-with-name-catlm-runnable-overall}{36.64}
\DefMacro{res-user-view-with-name-catlm-ne2e-with-name-catlm-runnable-overall-avg}{36.64}
\DefMacro{res-user-view-with-name-catlm-ne2e-with-name-catlm-runnable-overall-max}{36.64}
\DefMacro{res-user-view-with-name-catlm-ne2e-with-name-catlm-runnable-overall-min}{36.64}
\DefMacro{res-user-view-with-name-catlm-ne2e-with-name-catlm-timeout-avg}{0.50}
\DefMacro{res-user-view-with-name-catlm-ne2e-with-name-catlm-timeout-max}{0.50}
\DefMacro{res-user-view-with-name-catlm-ne2e-with-name-catlm-timeout-min}{0.50}
\DefMacro{res-user-view-with-name-selected-434-conditionnestack2e-with-name-zero-shot-lora-codellama-7b-bleu}{38.23}
\DefMacro{res-user-view-with-name-selected-434-conditionnestack2e-with-name-zero-shot-lora-codellama-7b-bleu-avg}{38.23}
\DefMacro{res-user-view-with-name-selected-434-conditionnestack2e-with-name-zero-shot-lora-codellama-7b-bleu-max}{38.23}
\DefMacro{res-user-view-with-name-selected-434-conditionnestack2e-with-name-zero-shot-lora-codellama-7b-bleu-min}{38.23}
\DefMacro{res-user-view-with-name-selected-434-conditionnestack2e-with-name-zero-shot-lora-codellama-7b-code-bleu}{48.63}
\DefMacro{res-user-view-with-name-selected-434-conditionnestack2e-with-name-zero-shot-lora-codellama-7b-code-bleu-avg}{48.63}
\DefMacro{res-user-view-with-name-selected-434-conditionnestack2e-with-name-zero-shot-lora-codellama-7b-code-bleu-max}{48.63}
\DefMacro{res-user-view-with-name-selected-434-conditionnestack2e-with-name-zero-shot-lora-codellama-7b-code-bleu-min}{48.63}
\DefMacro{res-user-view-with-name-selected-434-conditionnestack2e-with-name-zero-shot-lora-codellama-7b-edit-sim}{66.96}
\DefMacro{res-user-view-with-name-selected-434-conditionnestack2e-with-name-zero-shot-lora-codellama-7b-edit-sim-avg}{66.96}
\DefMacro{res-user-view-with-name-selected-434-conditionnestack2e-with-name-zero-shot-lora-codellama-7b-edit-sim-max}{66.96}
\DefMacro{res-user-view-with-name-selected-434-conditionnestack2e-with-name-zero-shot-lora-codellama-7b-edit-sim-min}{66.96}
\DefMacro{res-user-view-with-name-selected-434-conditionnestack2e-with-name-zero-shot-lora-codellama-7b-rouge-f}{61.28}
\DefMacro{res-user-view-with-name-selected-434-conditionnestack2e-with-name-zero-shot-lora-codellama-7b-rouge-f-avg}{61.28}
\DefMacro{res-user-view-with-name-selected-434-conditionnestack2e-with-name-zero-shot-lora-codellama-7b-rouge-f-max}{61.28}
\DefMacro{res-user-view-with-name-selected-434-conditionnestack2e-with-name-zero-shot-lora-codellama-7b-rouge-f-min}{61.28}
\DefMacro{res-user-view-with-name-selected-434-conditionnestack2e-with-name-zero-shot-lora-codellama-7b-rouge-p}{56.39}
\DefMacro{res-user-view-with-name-selected-434-conditionnestack2e-with-name-zero-shot-lora-codellama-7b-rouge-p-avg}{56.39}
\DefMacro{res-user-view-with-name-selected-434-conditionnestack2e-with-name-zero-shot-lora-codellama-7b-rouge-p-max}{56.39}
\DefMacro{res-user-view-with-name-selected-434-conditionnestack2e-with-name-zero-shot-lora-codellama-7b-rouge-p-min}{56.39}
\DefMacro{res-user-view-with-name-selected-434-conditionnestack2e-with-name-zero-shot-lora-codellama-7b-rouge-r}{71.47}
\DefMacro{res-user-view-with-name-selected-434-conditionnestack2e-with-name-zero-shot-lora-codellama-7b-rouge-r-avg}{71.47}
\DefMacro{res-user-view-with-name-selected-434-conditionnestack2e-with-name-zero-shot-lora-codellama-7b-rouge-r-max}{71.47}
\DefMacro{res-user-view-with-name-selected-434-conditionnestack2e-with-name-zero-shot-lora-codellama-7b-rouge-r-min}{71.47}
\DefMacro{res-user-view-with-name-selected-434-conditionnestack2e-with-name-zero-shot-lora-codellama-7b-xmatch}{5.53}
\DefMacro{res-user-view-with-name-selected-434-conditionnestack2e-with-name-zero-shot-lora-codellama-7b-xmatch-avg}{5.53}
\DefMacro{res-user-view-with-name-selected-434-conditionnestack2e-with-name-zero-shot-lora-codellama-7b-xmatch-max}{5.53}
\DefMacro{res-user-view-with-name-selected-434-conditionnestack2e-with-name-zero-shot-lora-codellama-7b-xmatch-min}{5.53}
\DefMacro{res-user-view-with-name-selected-434-conditionnestack2e-with-name-zero-shot-lora-codellama-7b-xmatch-top1}{5.53}
\DefMacro{res-user-view-with-name-selected-434-conditionnestack2e-with-name-zero-shot-lora-codellama-7b-compilable-avg}{57.30}
\DefMacro{res-user-view-with-name-selected-434-conditionnestack2e-with-name-zero-shot-lora-codellama-7b-compilable-max}{57.30}
\DefMacro{res-user-view-with-name-selected-434-conditionnestack2e-with-name-zero-shot-lora-codellama-7b-compilable-min}{57.30}
\DefMacro{res-user-view-with-name-selected-434-conditionnestack2e-with-name-zero-shot-lora-codellama-7b-coverage-avg}{33.87}
\DefMacro{res-user-view-with-name-selected-434-conditionnestack2e-with-name-zero-shot-lora-codellama-7b-coverage-max}{33.87}
\DefMacro{res-user-view-with-name-selected-434-conditionnestack2e-with-name-zero-shot-lora-codellama-7b-coverage-min}{33.87}
\DefMacro{res-user-view-with-name-selected-434-conditionnestack2e-with-name-zero-shot-lora-codellama-7b-match-avg}{96.02}
\DefMacro{res-user-view-with-name-selected-434-conditionnestack2e-with-name-zero-shot-lora-codellama-7b-match-max}{96.02}
\DefMacro{res-user-view-with-name-selected-434-conditionnestack2e-with-name-zero-shot-lora-codellama-7b-match-min}{96.02}
\DefMacro{res-user-view-with-name-selected-434-conditionnestack2e-with-name-zero-shot-lora-codellama-7b-runnable-avg}{73.06}
\DefMacro{res-user-view-with-name-selected-434-conditionnestack2e-with-name-zero-shot-lora-codellama-7b-runnable-max}{73.06}
\DefMacro{res-user-view-with-name-selected-434-conditionnestack2e-with-name-zero-shot-lora-codellama-7b-runnable-min}{73.06}
\DefMacro{res-user-view-with-name-selected-434-conditionnestack2e-with-name-zero-shot-lora-codellama-7b-runnable-overall}{40.17}
\DefMacro{res-user-view-with-name-selected-434-conditionnestack2e-with-name-zero-shot-lora-codellama-7b-runnable-overall-avg}{40.17}
\DefMacro{res-user-view-with-name-selected-434-conditionnestack2e-with-name-zero-shot-lora-codellama-7b-runnable-overall-max}{40.17}
\DefMacro{res-user-view-with-name-selected-434-conditionnestack2e-with-name-zero-shot-lora-codellama-7b-runnable-overall-min}{40.17}
\DefMacro{res-user-view-with-name-selected-434-conditionnestack2e-with-name-zero-shot-lora-codellama-7b-timeout-avg}{0.14}
\DefMacro{res-user-view-with-name-selected-434-conditionnestack2e-with-name-zero-shot-lora-codellama-7b-timeout-max}{0.14}
\DefMacro{res-user-view-with-name-selected-434-conditionnestack2e-with-name-zero-shot-lora-codellama-7b-timeout-min}{0.14}
\DefMacro{res-user-view-with-name-conditionnestack2e-with-name-ft-lora-codellama-7b-bleu}{63.13}
\DefMacro{res-user-view-with-name-conditionnestack2e-with-name-ft-lora-codellama-7b-bleu-avg}{63.13}
\DefMacro{res-user-view-with-name-conditionnestack2e-with-name-ft-lora-codellama-7b-bleu-max}{63.13}
\DefMacro{res-user-view-with-name-conditionnestack2e-with-name-ft-lora-codellama-7b-bleu-min}{63.13}
\DefMacro{res-user-view-with-name-conditionnestack2e-with-name-ft-lora-codellama-7b-code-bleu}{67.49}
\DefMacro{res-user-view-with-name-conditionnestack2e-with-name-ft-lora-codellama-7b-code-bleu-avg}{67.49}
\DefMacro{res-user-view-with-name-conditionnestack2e-with-name-ft-lora-codellama-7b-code-bleu-max}{67.49}
\DefMacro{res-user-view-with-name-conditionnestack2e-with-name-ft-lora-codellama-7b-code-bleu-min}{67.49}
\DefMacro{res-user-view-with-name-conditionnestack2e-with-name-ft-lora-codellama-7b-edit-sim}{85.32}
\DefMacro{res-user-view-with-name-conditionnestack2e-with-name-ft-lora-codellama-7b-edit-sim-avg}{85.32}
\DefMacro{res-user-view-with-name-conditionnestack2e-with-name-ft-lora-codellama-7b-edit-sim-max}{85.32}
\DefMacro{res-user-view-with-name-conditionnestack2e-with-name-ft-lora-codellama-7b-edit-sim-min}{85.32}
\DefMacro{res-user-view-with-name-conditionnestack2e-with-name-ft-lora-codellama-7b-rouge-f}{80.75}
\DefMacro{res-user-view-with-name-conditionnestack2e-with-name-ft-lora-codellama-7b-rouge-f-avg}{80.75}
\DefMacro{res-user-view-with-name-conditionnestack2e-with-name-ft-lora-codellama-7b-rouge-f-max}{80.75}
\DefMacro{res-user-view-with-name-conditionnestack2e-with-name-ft-lora-codellama-7b-rouge-f-min}{80.75}
\DefMacro{res-user-view-with-name-conditionnestack2e-with-name-ft-lora-codellama-7b-rouge-p}{82.85}
\DefMacro{res-user-view-with-name-conditionnestack2e-with-name-ft-lora-codellama-7b-rouge-p-avg}{82.85}
\DefMacro{res-user-view-with-name-conditionnestack2e-with-name-ft-lora-codellama-7b-rouge-p-max}{82.85}
\DefMacro{res-user-view-with-name-conditionnestack2e-with-name-ft-lora-codellama-7b-rouge-p-min}{82.85}
\DefMacro{res-user-view-with-name-conditionnestack2e-with-name-ft-lora-codellama-7b-rouge-r}{80.81}
\DefMacro{res-user-view-with-name-conditionnestack2e-with-name-ft-lora-codellama-7b-rouge-r-avg}{80.81}
\DefMacro{res-user-view-with-name-conditionnestack2e-with-name-ft-lora-codellama-7b-rouge-r-max}{80.81}
\DefMacro{res-user-view-with-name-conditionnestack2e-with-name-ft-lora-codellama-7b-rouge-r-min}{80.81}
\DefMacro{res-user-view-with-name-conditionnestack2e-with-name-ft-lora-codellama-7b-xmatch}{19.05}
\DefMacro{res-user-view-with-name-conditionnestack2e-with-name-ft-lora-codellama-7b-xmatch-avg}{19.05}
\DefMacro{res-user-view-with-name-conditionnestack2e-with-name-ft-lora-codellama-7b-xmatch-max}{19.05}
\DefMacro{res-user-view-with-name-conditionnestack2e-with-name-ft-lora-codellama-7b-xmatch-min}{19.05}
\DefMacro{res-user-view-with-name-conditionnestack2e-with-name-ft-lora-codellama-7b-xmatch-top1}{19.05}
\DefMacro{res-user-view-with-name-conditionnestack2e-with-name-ft-lora-codellama-7b-compilable-avg}{82.10}
\DefMacro{res-user-view-with-name-conditionnestack2e-with-name-ft-lora-codellama-7b-compilable-max}{82.10}
\DefMacro{res-user-view-with-name-conditionnestack2e-with-name-ft-lora-codellama-7b-compilable-min}{82.10}
\DefMacro{res-user-view-with-name-conditionnestack2e-with-name-ft-lora-codellama-7b-coverage-avg}{59.45}
\DefMacro{res-user-view-with-name-conditionnestack2e-with-name-ft-lora-codellama-7b-coverage-max}{59.45}
\DefMacro{res-user-view-with-name-conditionnestack2e-with-name-ft-lora-codellama-7b-coverage-min}{59.45}
\DefMacro{res-user-view-with-name-conditionnestack2e-with-name-ft-lora-codellama-7b-match-avg}{100.00}
\DefMacro{res-user-view-with-name-conditionnestack2e-with-name-ft-lora-codellama-7b-match-max}{100.00}
\DefMacro{res-user-view-with-name-conditionnestack2e-with-name-ft-lora-codellama-7b-match-min}{100.00}
\DefMacro{res-user-view-with-name-conditionnestack2e-with-name-ft-lora-codellama-7b-runnable-avg}{82.05}
\DefMacro{res-user-view-with-name-conditionnestack2e-with-name-ft-lora-codellama-7b-runnable-max}{82.05}
\DefMacro{res-user-view-with-name-conditionnestack2e-with-name-ft-lora-codellama-7b-runnable-min}{82.05}
\DefMacro{res-user-view-with-name-conditionnestack2e-with-name-ft-lora-codellama-7b-runnable-overall}{67.36}
\DefMacro{res-user-view-with-name-conditionnestack2e-with-name-ft-lora-codellama-7b-runnable-overall-avg}{67.36}
\DefMacro{res-user-view-with-name-conditionnestack2e-with-name-ft-lora-codellama-7b-runnable-overall-max}{67.36}
\DefMacro{res-user-view-with-name-conditionnestack2e-with-name-ft-lora-codellama-7b-runnable-overall-min}{67.36}
\DefMacro{res-user-view-with-name-conditionnestack2e-with-name-ft-lora-codellama-7b-timeout-avg}{0.65}
\DefMacro{res-user-view-with-name-conditionnestack2e-with-name-ft-lora-codellama-7b-timeout-max}{0.65}
\DefMacro{res-user-view-with-name-conditionnestack2e-with-name-ft-lora-codellama-7b-timeout-min}{0.65}
\DefMacro{res-user-view-with-name-conditionnestack2e-all-with-name-ft-lora-codellama-7b-bleu}{62.10}
\DefMacro{res-user-view-with-name-conditionnestack2e-all-with-name-ft-lora-codellama-7b-bleu-avg}{61.76}
\DefMacro{res-user-view-with-name-conditionnestack2e-all-with-name-ft-lora-codellama-7b-bleu-max}{70.01}
\DefMacro{res-user-view-with-name-conditionnestack2e-all-with-name-ft-lora-codellama-7b-bleu-min}{53.80}
\DefMacro{res-user-view-with-name-conditionnestack2e-all-with-name-ft-lora-codellama-7b-code-bleu}{66.52}
\DefMacro{res-user-view-with-name-conditionnestack2e-all-with-name-ft-lora-codellama-7b-code-bleu-avg}{66.64}
\DefMacro{res-user-view-with-name-conditionnestack2e-all-with-name-ft-lora-codellama-7b-code-bleu-max}{74.08}
\DefMacro{res-user-view-with-name-conditionnestack2e-all-with-name-ft-lora-codellama-7b-code-bleu-min}{59.92}
\DefMacro{res-user-view-with-name-conditionnestack2e-all-with-name-ft-lora-codellama-7b-edit-sim}{85.20}
\DefMacro{res-user-view-with-name-conditionnestack2e-all-with-name-ft-lora-codellama-7b-edit-sim-avg}{85.26}
\DefMacro{res-user-view-with-name-conditionnestack2e-all-with-name-ft-lora-codellama-7b-edit-sim-max}{90.09}
\DefMacro{res-user-view-with-name-conditionnestack2e-all-with-name-ft-lora-codellama-7b-edit-sim-min}{79.88}
\DefMacro{res-user-view-with-name-conditionnestack2e-all-with-name-ft-lora-codellama-7b-rouge-f}{80.08}
\DefMacro{res-user-view-with-name-conditionnestack2e-all-with-name-ft-lora-codellama-7b-rouge-f-avg}{80.02}
\DefMacro{res-user-view-with-name-conditionnestack2e-all-with-name-ft-lora-codellama-7b-rouge-f-max}{85.50}
\DefMacro{res-user-view-with-name-conditionnestack2e-all-with-name-ft-lora-codellama-7b-rouge-f-min}{74.19}
\DefMacro{res-user-view-with-name-conditionnestack2e-all-with-name-ft-lora-codellama-7b-rouge-p}{82.02}
\DefMacro{res-user-view-with-name-conditionnestack2e-all-with-name-ft-lora-codellama-7b-rouge-p-avg}{81.93}
\DefMacro{res-user-view-with-name-conditionnestack2e-all-with-name-ft-lora-codellama-7b-rouge-p-max}{88.06}
\DefMacro{res-user-view-with-name-conditionnestack2e-all-with-name-ft-lora-codellama-7b-rouge-p-min}{74.57}
\DefMacro{res-user-view-with-name-conditionnestack2e-all-with-name-ft-lora-codellama-7b-rouge-r}{80.14}
\DefMacro{res-user-view-with-name-conditionnestack2e-all-with-name-ft-lora-codellama-7b-rouge-r-avg}{80.15}
\DefMacro{res-user-view-with-name-conditionnestack2e-all-with-name-ft-lora-codellama-7b-rouge-r-max}{85.20}
\DefMacro{res-user-view-with-name-conditionnestack2e-all-with-name-ft-lora-codellama-7b-rouge-r-min}{75.28}
\DefMacro{res-user-view-with-name-conditionnestack2e-all-with-name-ft-lora-codellama-7b-xmatch}{15.28}
\DefMacro{res-user-view-with-name-conditionnestack2e-all-with-name-ft-lora-codellama-7b-xmatch-avg}{15.03}
\DefMacro{res-user-view-with-name-conditionnestack2e-all-with-name-ft-lora-codellama-7b-xmatch-max}{22.92}
\DefMacro{res-user-view-with-name-conditionnestack2e-all-with-name-ft-lora-codellama-7b-xmatch-min}{9.62}
\DefMacro{res-user-view-with-name-conditionnestack2e-all-with-name-ft-lora-codellama-7b-xmatch-top1}{15.28}
\DefMacro{res-user-view-with-name-conditionnestack2e-all-with-name-ft-lora-codellama-7b-compilable-avg}{82.93}
\DefMacro{res-user-view-with-name-conditionnestack2e-all-with-name-ft-lora-codellama-7b-compilable-max}{93.54}
\DefMacro{res-user-view-with-name-conditionnestack2e-all-with-name-ft-lora-codellama-7b-compilable-min}{69.30}
\DefMacro{res-user-view-with-name-conditionnestack2e-all-with-name-ft-lora-codellama-7b-coverage-avg}{55.78}
\DefMacro{res-user-view-with-name-conditionnestack2e-all-with-name-ft-lora-codellama-7b-coverage-max}{71.28}
\DefMacro{res-user-view-with-name-conditionnestack2e-all-with-name-ft-lora-codellama-7b-coverage-min}{39.26}
\DefMacro{res-user-view-with-name-conditionnestack2e-all-with-name-ft-lora-codellama-7b-match-avg}{100.00}
\DefMacro{res-user-view-with-name-conditionnestack2e-all-with-name-ft-lora-codellama-7b-match-max}{100.00}
\DefMacro{res-user-view-with-name-conditionnestack2e-all-with-name-ft-lora-codellama-7b-match-min}{100.00}
\DefMacro{res-user-view-with-name-conditionnestack2e-all-with-name-ft-lora-codellama-7b-runnable-avg}{77.04}
\DefMacro{res-user-view-with-name-conditionnestack2e-all-with-name-ft-lora-codellama-7b-runnable-max}{86.91}
\DefMacro{res-user-view-with-name-conditionnestack2e-all-with-name-ft-lora-codellama-7b-runnable-min}{65.21}
\DefMacro{res-user-view-with-name-conditionnestack2e-all-with-name-ft-lora-codellama-7b-runnable-overall}{81.29}
\DefMacro{res-user-view-with-name-conditionnestack2e-all-with-name-ft-lora-codellama-7b-runnable-overall-avg}{64.94}
\DefMacro{res-user-view-with-name-conditionnestack2e-all-with-name-ft-lora-codellama-7b-runnable-overall-max}{81.29}
\DefMacro{res-user-view-with-name-conditionnestack2e-all-with-name-ft-lora-codellama-7b-runnable-overall-min}{46.11}
\DefMacro{res-user-view-with-name-conditionnestack2e-all-with-name-ft-lora-codellama-7b-timeout-avg}{0.52}
\DefMacro{res-user-view-with-name-conditionnestack2e-all-with-name-ft-lora-codellama-7b-timeout-max}{0.56}
\DefMacro{res-user-view-with-name-conditionnestack2e-all-with-name-ft-lora-codellama-7b-timeout-min}{0.42}

%% file: main.bbl

\begin{thebibliography}{44}


\ifx \showCODEN    \undefined \def \showCODEN     #1{\unskip}     \fi
\ifx \showDOI      \undefined \def \showDOI       #1{#1}\fi
\ifx \showISBNx    \undefined \def \showISBNx     #1{\unskip}     \fi
\ifx \showISBNxiii \undefined \def \showISBNxiii  #1{\unskip}     \fi
\ifx \showISSN     \undefined \def \showISSN      #1{\unskip}     \fi
\ifx \showLCCN     \undefined \def \showLCCN      #1{\unskip}     \fi
\ifx \shownote     \undefined \def \shownote      #1{#1}          \fi
\ifx \showarticletitle \undefined \def \showarticletitle #1{#1}   \fi
\ifx \showURL      \undefined \def \showURL       {\relax}        \fi
\providecommand\bibfield[2]{#2}
\providecommand\bibinfo[2]{#2}
\providecommand\natexlab[1]{#1}
\providecommand\showeprint[2][]{arXiv:#2}

\bibitem[Almulla and Gay(2020)]%
        {AlmullaETAL20GeneratingExceptionTriggeringTests}
\bibfield{author}{\bibinfo{person}{Hussein Almulla} {and}
  \bibinfo{person}{Gregory Gay}.} \bibinfo{year}{2020}\natexlab{}.
\newblock \showarticletitle{Learning how to search: Generating
  exception-triggering tests through adaptive fitness function selection}. In
  \bibinfo{booktitle}{\emph{International Conference on Software Testing,
  Verification, and Validation}}. \bibinfo{pages}{63--73}.
\newblock


\bibitem[Bernardo et~al\mbox{.}(2011)]%
        {BernardoETAL11AgileTestingOfExceptionalBehavior}
\bibfield{author}{\bibinfo{person}{Rafael~Di Bernardo},
  \bibinfo{person}{Ricardo Sales~Jr.}, \bibinfo{person}{Fernando Castor},
  \bibinfo{person}{Roberta Coelho}, \bibinfo{person}{Nelio Cacho}, {and}
  \bibinfo{person}{Sergio Soares}.} \bibinfo{year}{2011}\natexlab{}.
\newblock \showarticletitle{Agile testing of exceptional behavior}. In
  \bibinfo{booktitle}{\emph{Brazilian Symposium on Software Engineering}}.
  \bibinfo{pages}{204--213}.
\newblock


\bibitem[Blasi et~al\mbox{.}(2022)]%
        {blasi2022call}
\bibfield{author}{\bibinfo{person}{Arianna Blasi}, \bibinfo{person}{Alessandra
  Gorla}, \bibinfo{person}{Michael~D Ernst}, {and} \bibinfo{person}{Mauro
  Pezz{\`e}}.} \bibinfo{year}{2022}\natexlab{}.
\newblock \showarticletitle{Call me maybe: Using {NLP} to automatically
  generate unit test cases respecting temporal constraints}. In
  \bibinfo{booktitle}{\emph{Automated Software Engineering}}.
  \bibinfo{pages}{1--11}.
\newblock


\bibitem[Bouzenia and Pradel(2023)]%
        {bouzenia2023say}
\bibfield{author}{\bibinfo{person}{Islem Bouzenia} {and}
  \bibinfo{person}{Michael Pradel}.} \bibinfo{year}{2023}\natexlab{}.
\newblock \showarticletitle{When to say what: Learning to find
  condition-message inconsistencies}. In
  \bibinfo{booktitle}{\emph{International Conference on Software Engineering}}.
  \bibinfo{pages}{868--880}.
\newblock


\bibitem[{Conda}(2024)]%
        {Conda}
\bibfield{author}{\bibinfo{person}{{Conda}}.} \bibinfo{year}{2024}\natexlab{}.
\newblock \bibinfo{title}{{Conda}}.
\newblock
  \bibinfo{howpublished}{\url{https://docs.conda.io/projects/conda/en/stable}}.
\newblock


\bibitem[Daka(2018)]%
        {daka2018improving}
\bibfield{author}{\bibinfo{person}{Ermira Daka}.}
  \bibinfo{year}{2018}\natexlab{}.
\newblock \emph{\bibinfo{title}{Improving readability in automatic unit test
  generation}}.
\newblock \bibinfo{thesistype}{Ph.\,D. Dissertation}.
  \bibinfo{school}{University of Sheffield}.
\newblock


\bibitem[Daka et~al\mbox{.}(2017)]%
        {daka2017generating}
\bibfield{author}{\bibinfo{person}{Ermira Daka},
  \bibinfo{person}{Jos{\'e}~Miguel Rojas}, {and} \bibinfo{person}{Gordon
  Fraser}.} \bibinfo{year}{2017}\natexlab{}.
\newblock \showarticletitle{Generating unit tests with descriptive names or:
  {W}ould you name your children thing1 and thing2?}. In
  \bibinfo{booktitle}{\emph{International Symposium on Software Testing and
  Analysis}}. \bibinfo{pages}{57--67}.
\newblock


\bibitem[Dalton et~al\mbox{.}(2020)]%
        {DaltonETAL20ExceptionalBehaviorTesting}
\bibfield{author}{\bibinfo{person}{Francisco Dalton},
  \bibinfo{person}{M\'{a}rcio Ribeiro}, \bibinfo{person}{Gustavo Pinto},
  \bibinfo{person}{Leo Fernandes}, \bibinfo{person}{Rohit Gheyi}, {and}
  \bibinfo{person}{Baldoino Fonseca}.} \bibinfo{year}{2020}\natexlab{}.
\newblock \showarticletitle{Is exceptional behavior testing an exception? {A}n
  empirical assessment using {J}ava automated tests}. In
  \bibinfo{booktitle}{\emph{International Conference on Evaluation and
  Assessment in Software Engineering}}. \bibinfo{pages}{170--179}.
\newblock


\bibitem[Dettmers et~al\mbox{.}(2024)]%
        {dettmers2024int8}
\bibfield{author}{\bibinfo{person}{Tim Dettmers}, \bibinfo{person}{Mike Lewis},
  \bibinfo{person}{Younes Belkada}, {and} \bibinfo{person}{Luke Zettlemoyer}.}
  \bibinfo{year}{2024}\natexlab{}.
\newblock \showarticletitle{LLM.int8(): 8-bit matrix multiplication for
  transformers at scale}. In \bibinfo{booktitle}{\emph{International Conference
  on Neural Information Processing Systems}}.
\newblock


\bibitem[El~Haji et~al\mbox{.}(2024)]%
        {el2024using}
\bibfield{author}{\bibinfo{person}{Khalid El~Haji}, \bibinfo{person}{Carolin
  Brandt}, {and} \bibinfo{person}{Andy Zaidman}.}
  \bibinfo{year}{2024}\natexlab{}.
\newblock \showarticletitle{Using {G}itHub {C}opilot for test generation in
  {P}ython: {A}n empirical study}.
\newblock \bibinfo{journal}{\emph{International Workshop on Automation of
  Software Test}}, \bibinfo{pages}{45--55}.
\newblock


\bibitem[Ernst et~al\mbox{.}(2007)]%
        {ernst2007daikon}
\bibfield{author}{\bibinfo{person}{Michael~D Ernst}, \bibinfo{person}{Jeff~H
  Perkins}, \bibinfo{person}{Philip~J Guo}, \bibinfo{person}{Stephen McCamant},
  \bibinfo{person}{Carlos Pacheco}, \bibinfo{person}{Matthew~S Tschantz}, {and}
  \bibinfo{person}{Chen Xiao}.} \bibinfo{year}{2007}\natexlab{}.
\newblock \showarticletitle{The {D}aikon system for dynamic detection of likely
  invariants}.
\newblock \bibinfo{journal}{\emph{Science of computer programming}}
  \bibinfo{volume}{69}, \bibinfo{number}{1-3}, \bibinfo{pages}{35--45}.
\newblock


\bibitem[Fraser and Arcuri(2011)]%
        {FraserAndArcuri11EvoSuite}
\bibfield{author}{\bibinfo{person}{Gordon Fraser} {and} \bibinfo{person}{Andrea
  Arcuri}.} \bibinfo{year}{2011}\natexlab{}.
\newblock \showarticletitle{{EvoSuite}: Automatic test suite generation for
  object-oriented software}. In \bibinfo{booktitle}{\emph{International
  Symposium on the Foundations of Software Engineering}}.
  \bibinfo{pages}{416--419}.
\newblock


\bibitem[Godefroid(2012)]%
        {godefroid2012test}
\bibfield{author}{\bibinfo{person}{Patrice Godefroid}.}
  \bibinfo{year}{2012}\natexlab{}.
\newblock \showarticletitle{Test generation using symbolic execution}. In
  \bibinfo{booktitle}{\emph{Annual Conference on Foundations of Software
  Technology and Theoretical Computer Science}}.
\newblock


\bibitem[Goffi et~al\mbox{.}(2016)]%
        {GoffiETAL16AutomaticGenerationOfOraclesForExceptionalBehaviors}
\bibfield{author}{\bibinfo{person}{Alberto Goffi}, \bibinfo{person}{Alessandra
  Gorla}, \bibinfo{person}{Michael~D. Ernst}, {and} \bibinfo{person}{Mauro
  Pezz\`{e}}.} \bibinfo{year}{2016}\natexlab{}.
\newblock \showarticletitle{Automatic generation of oracles for exceptional
  behaviors}. In \bibinfo{booktitle}{\emph{International Symposium on Software
  Testing and Analysis}}. \bibinfo{pages}{213--224}.
\newblock


\bibitem[Gosling(2000)]%
        {gosling2000java}
\bibfield{author}{\bibinfo{person}{James Gosling}.}
  \bibinfo{year}{2000}\natexlab{}.
\newblock \bibinfo{booktitle}{\emph{The Java language specification}}.
\newblock \bibinfo{publisher}{Addison-Wesley Professional}.
\newblock


\bibitem[Harman and McMinn(2009)]%
        {harman2009theoretical}
\bibfield{author}{\bibinfo{person}{Mark Harman} {and} \bibinfo{person}{Phil
  McMinn}.} \bibinfo{year}{2009}\natexlab{}.
\newblock \showarticletitle{A theoretical and empirical study of search-based
  testing: Local, global, and hybrid search}.
\newblock \bibinfo{journal}{\emph{Transactions on Software Engineering}}
  \bibinfo{volume}{36}, \bibinfo{number}{2}, \bibinfo{pages}{226--247}.
\newblock


\bibitem[Hejlsberg et~al\mbox{.}(2003)]%
        {hejlsberg2003c}
\bibfield{author}{\bibinfo{person}{Anders Hejlsberg}, \bibinfo{person}{Scott
  Wiltamuth}, {and} \bibinfo{person}{Peter Golde}.}
  \bibinfo{year}{2003}\natexlab{}.
\newblock \bibinfo{booktitle}{\emph{C\# language specification}}.
\newblock \bibinfo{publisher}{Addison-Wesley Longman Publishing Co., Inc.}
\newblock


\bibitem[Holmes et~al\mbox{.}(2020)]%
        {HolmesETAL20TestGeneration}
\bibfield{author}{\bibinfo{person}{Josie Holmes}, \bibinfo{person}{Iftekhar
  Ahmed}, \bibinfo{person}{Caius Brindescu}, \bibinfo{person}{Rahul Gopinath},
  \bibinfo{person}{He Zhang}, {and} \bibinfo{person}{Alex Groce}.}
  \bibinfo{year}{2020}\natexlab{}.
\newblock \showarticletitle{Using relative lines of code to guide automated
  test generation for {Python}}.
\newblock \bibinfo{journal}{\emph{Transactions on Software Engineering and
  Methodology}} \bibinfo{volume}{29}, \bibinfo{number}{4},
  \bibinfo{pages}{1--38}.
\newblock


\bibitem[Husain et~al\mbox{.}(2019)]%
        {CodeSearchNet}
\bibfield{author}{\bibinfo{person}{Hamel Husain}, \bibinfo{person}{Ho-Hsiang
  Wu}, \bibinfo{person}{Tiferet Gazit}, \bibinfo{person}{Miltiadis Allamanis},
  {and} \bibinfo{person}{Marc Brockschmidt}.} \bibinfo{year}{2019}\natexlab{}.
\newblock \showarticletitle{{C}ode{S}earch{N}et challenge: {E}valuating the
  state of semantic code search}.
\newblock \bibinfo{journal}{\emph{arXiv preprint arXiv:1909.09436}}.
\newblock


\bibitem[Lemieux et~al\mbox{.}(2023)]%
        {LemieuxETAL23CodaMosa}
\bibfield{author}{\bibinfo{person}{Caroline Lemieux},
  \bibinfo{person}{Jeevana~Priya Inala}, \bibinfo{person}{Shuvendu~K. Lahiri},
  {and} \bibinfo{person}{Siddhartha Sen}.} \bibinfo{year}{2023}\natexlab{}.
\newblock \showarticletitle{{CodaMosa}: Escaping coverage plateaus in test
  generation with pre-trained large language models}. In
  \bibinfo{booktitle}{\emph{International Conference on Software Engineering}}.
  \bibinfo{pages}{919--931}.
\newblock


\bibitem[Lima et~al\mbox{.}(2021)]%
        {LimaETAL21AssessingExceptionHandlingTesting}
\bibfield{author}{\bibinfo{person}{Luan~P. Lima}, \bibinfo{person}{Lincoln~S.
  Rocha}, \bibinfo{person}{Carla I.~M. Bezerra}, {and} \bibinfo{person}{Matheus
  Paixao}.} \bibinfo{year}{2021}\natexlab{}.
\newblock \showarticletitle{Assessing exception handling testing practices in
  open-source libraries}.
\newblock \bibinfo{journal}{\emph{Empirical Software Engineering}}
  \bibinfo{volume}{26}, \bibinfo{number}{5}.
\newblock


\bibitem[Liu et~al\mbox{.}(2023)]%
        {LiuISSTA23EXLI}
\bibfield{author}{\bibinfo{person}{Yu Liu}, \bibinfo{person}{Pengyu Nie},
  \bibinfo{person}{Anna Guo}, \bibinfo{person}{Milos Gligoric}, {and}
  \bibinfo{person}{Owolabi Legunsen}.} \bibinfo{year}{2023}\natexlab{}.
\newblock \showarticletitle{Extracting Inline Tests from Unit Tests}. In
  \bibinfo{booktitle}{\emph{International Symposium on Software Testing and
  Analysis}}. \bibinfo{pages}{1--13}.
\newblock


\bibitem[Liu et~al\mbox{.}(2024)]%
        {LiuFSE24EXLI}
\bibfield{author}{\bibinfo{person}{Yu Liu}, \bibinfo{person}{Aditya Thimmaiah},
  \bibinfo{person}{Owolabi Legunsen}, {and} \bibinfo{person}{Milos Gligoric}.}
  \bibinfo{year}{2024}\natexlab{}.
\newblock \showarticletitle{{E}x{L}i: An Inline-Test Generation Tool for
  {J}ava}. In \bibinfo{booktitle}{\emph{International Symposium on Software
  Testing and Analysis}}. \bibinfo{pages}{1--5}.
\newblock


\bibitem[Marcilio and Furia(2021)]%
        {MarcilioFuria21HowJavaProgrammersTestExceptionalBehavior}
\bibfield{author}{\bibinfo{person}{Diego Marcilio} {and}
  \bibinfo{person}{Carlo~A. Furia}.} \bibinfo{year}{2021}\natexlab{}.
\newblock \showarticletitle{How {J}ava programmers test exceptional behavior}.
  In \bibinfo{booktitle}{\emph{International Working Conference on Mining
  Software Repositories}}. \bibinfo{pages}{207--218}.
\newblock


\bibitem[Nie(2023)]%
        {Nie23Thesis}
\bibfield{author}{\bibinfo{person}{Pengyu Nie}.}
  \bibinfo{year}{2023}\natexlab{}.
\newblock \emph{\bibinfo{title}{Machine learning for executable code in
  software testing and verification}}.
\newblock \bibinfo{thesistype}{Ph.\,D. Dissertation}. \bibinfo{school}{The
  University of Texas at Austin}.
\newblock


\bibitem[Nie et~al\mbox{.}(2023)]%
        {NieETAL23TeCo}
\bibfield{author}{\bibinfo{person}{Pengyu Nie}, \bibinfo{person}{Rahul
  Banerjee}, \bibinfo{person}{Junyi~Jessy Li}, \bibinfo{person}{Raymond~J.
  Mooney}, {and} \bibinfo{person}{Milos Gligoric}.}
  \bibinfo{year}{2023}\natexlab{}.
\newblock \showarticletitle{Learning deep semantics for test completion}. In
  \bibinfo{booktitle}{\emph{International Conference on Software Engineering}}.
  \bibinfo{pages}{2111--2123}.
\newblock


\bibitem[OpenAI(2024)]%
        {gpt3.5-turbo}
\bibfield{author}{\bibinfo{person}{OpenAI}.} \bibinfo{year}{2024}\natexlab{}.
\newblock \bibinfo{title}{{GPT}-3.5-turbo}.
\newblock
  \bibinfo{howpublished}{\url{https://platform.openai.com/docs/models/gpt-3-5-turbo}}.
\newblock


\bibitem[Pacheco et~al\mbox{.}(2007)]%
        {PachecoETAL07Randoop}
\bibfield{author}{\bibinfo{person}{Carlos Pacheco},
  \bibinfo{person}{Shuvendu~K. Lahiri}, \bibinfo{person}{Michael~D. Ernst},
  {and} \bibinfo{person}{Thomas Ball}.} \bibinfo{year}{2007}\natexlab{}.
\newblock \showarticletitle{Feedback-Directed random test generation}. In
  \bibinfo{booktitle}{\emph{International Conference on Software Engineering}}.
  \bibinfo{pages}{75--84}.
\newblock


\bibitem[Panichella et~al\mbox{.}(2022)]%
        {panichella2022test}
\bibfield{author}{\bibinfo{person}{Annibale Panichella},
  \bibinfo{person}{Sebastiano Panichella}, \bibinfo{person}{Gordon Fraser},
  \bibinfo{person}{Anand~Ashok Sawant}, {and} \bibinfo{person}{Vincent~J
  Hellendoorn}.} \bibinfo{year}{2022}\natexlab{}.
\newblock \showarticletitle{Test smells 20 years later: Detectability,
  validity, and reliability}.
\newblock \bibinfo{journal}{\emph{Empirical Software Engineering}}
  \bibinfo{volume}{27}, \bibinfo{number}{7}, \bibinfo{pages}{170}.
\newblock


\bibitem[Rao et~al\mbox{.}(2023)]%
        {RaoETAL23CAT}
\bibfield{author}{\bibinfo{person}{Nikitha Rao}, \bibinfo{person}{Kush Jain},
  \bibinfo{person}{Uri Alon}, \bibinfo{person}{Claire~Le Goues}, {and}
  \bibinfo{person}{Vincent~J. Hellendoorn}.} \bibinfo{year}{2023}\natexlab{}.
\newblock \showarticletitle{{CAT-LM}: Training language models on aligned code
  and tests}. In \bibinfo{booktitle}{\emph{Automated Software Engineering}}.
  \bibinfo{pages}{409--420}.
\newblock


\bibitem[Robinson et~al\mbox{.}(2011)]%
        {RobinsonETAL11ScalingTestGen}
\bibfield{author}{\bibinfo{person}{Brian Robinson}, \bibinfo{person}{Michael~D
  Ernst}, \bibinfo{person}{Jeff~H Perkins}, \bibinfo{person}{Vinay Augustine},
  {and} \bibinfo{person}{Nuo Li}.} \bibinfo{year}{2011}\natexlab{}.
\newblock \showarticletitle{Scaling up automated test generation: Automatically
  generating maintainable regression unit tests for programs}. In
  \bibinfo{booktitle}{\emph{Automated Software Engineering}}.
  \bibinfo{pages}{23--32}.
\newblock


\bibitem[Roziere et~al\mbox{.}(2023)]%
        {roziere2023code}
\bibfield{author}{\bibinfo{person}{Baptiste Roziere}, \bibinfo{person}{Jonas
  Gehring}, \bibinfo{person}{Fabian Gloeckle}, \bibinfo{person}{Sten Sootla},
  \bibinfo{person}{Itai Gat}, \bibinfo{person}{Xiaoqing~Ellen Tan},
  \bibinfo{person}{Yossi Adi}, \bibinfo{person}{Jingyu Liu},
  \bibinfo{person}{Tal Remez}, \bibinfo{person}{J{\'e}r{\'e}my Rapin},
  {et~al\mbox{.}}} \bibinfo{year}{2023}\natexlab{}.
\newblock \showarticletitle{Code {L}lama: Open foundation models for code}.
\newblock \bibinfo{journal}{\emph{arXiv preprint arXiv:2308.12950}}.
\newblock


\bibitem[Ryan et~al\mbox{.}(2024)]%
        {ryan2024code}
\bibfield{author}{\bibinfo{person}{Gabriel Ryan}, \bibinfo{person}{Siddhartha
  Jain}, \bibinfo{person}{Mingyue Shang}, \bibinfo{person}{Shiqi Wang},
  \bibinfo{person}{Xiaofei Ma}, \bibinfo{person}{Murali~Krishna Ramanathan},
  {and} \bibinfo{person}{Baishakhi Ray}.} \bibinfo{year}{2024}\natexlab{}.
\newblock \showarticletitle{Code-Aware prompting: A study of coverage guided
  test generation in regression setting using {LLM}}. In
  \bibinfo{booktitle}{\emph{International Symposium on the Foundations of
  Software Engineering}}. \bibinfo{pages}{951--971}.
\newblock


\bibitem[Sanh et~al\mbox{.}(2021)]%
        {sanh2021multitask}
\bibfield{author}{\bibinfo{person}{Victor Sanh}, \bibinfo{person}{Albert
  Webson}, \bibinfo{person}{Colin Raffel}, \bibinfo{person}{Stephen~H Bach},
  \bibinfo{person}{Lintang Sutawika}, \bibinfo{person}{Zaid Alyafeai},
  \bibinfo{person}{Antoine Chaffin}, \bibinfo{person}{Arnaud Stiegler},
  \bibinfo{person}{Teven~Le Scao}, \bibinfo{person}{Arun Raja},
  {et~al\mbox{.}}} \bibinfo{year}{2021}\natexlab{}.
\newblock \showarticletitle{Multitask prompted training enables zero-shot task
  generalization}.
\newblock \bibinfo{journal}{\emph{arXiv preprint arXiv:2110.08207}}.
\newblock


\bibitem[Sch{\"a}fer et~al\mbox{.}(2023)]%
        {SchaferETAL23TestPilot}
\bibfield{author}{\bibinfo{person}{Max Sch{\"a}fer}, \bibinfo{person}{Sarah
  Nadi}, \bibinfo{person}{Aryaz Eghbali}, {and} \bibinfo{person}{Frank Tip}.}
  \bibinfo{year}{2023}\natexlab{}.
\newblock \showarticletitle{An empirical evaluation of using large language
  models for automated unit test generation}.
\newblock \bibinfo{journal}{\emph{Transactions on Software Engineering}}.
\newblock


\bibitem[Tufano et~al\mbox{.}(2020)]%
        {TufanoETAL20TestGeneration}
\bibfield{author}{\bibinfo{person}{Michele Tufano}, \bibinfo{person}{Dawn
  Drain}, \bibinfo{person}{Alexey Svyatkovskiy}, \bibinfo{person}{Shao~Kun
  Deng}, {and} \bibinfo{person}{Neel Sundaresan}.}
  \bibinfo{year}{2020}\natexlab{}.
\newblock \showarticletitle{Unit test case generation with transformers and
  focal context}.
\newblock \bibinfo{journal}{\emph{arXiv preprint arXiv:2009.05617}}.
\newblock


\bibitem[VanRossum and Drake(2010)]%
        {vanrossum2010python}
\bibfield{author}{\bibinfo{person}{Guido VanRossum} {and}
  \bibinfo{person}{Fred~L Drake}.} \bibinfo{year}{2010}\natexlab{}.
\newblock \bibinfo{booktitle}{\emph{The {P}ython language reference}}.
  Vol.~\bibinfo{volume}{561}.
\newblock \bibinfo{publisher}{Python Software Foundation Amsterdam, The
  Netherlands}.
\newblock


\bibitem[Wang et~al\mbox{.}(2024)]%
        {wang2024software}
\bibfield{author}{\bibinfo{person}{Junjie Wang}, \bibinfo{person}{Yuchao
  Huang}, \bibinfo{person}{Chunyang Chen}, \bibinfo{person}{Zhe Liu},
  \bibinfo{person}{Song Wang}, {and} \bibinfo{person}{Qing Wang}.}
  \bibinfo{year}{2024}\natexlab{}.
\newblock \showarticletitle{Software testing with large language models:
  Survey, landscape, and vision}.
\newblock \bibinfo{journal}{\emph{Transactions on Software Engineering}}.
\newblock


\bibitem[Wang et~al\mbox{.}(2023)]%
        {selfInstruct}
\bibfield{author}{\bibinfo{person}{Yizhong Wang}, \bibinfo{person}{Yeganeh
  Kordi}, \bibinfo{person}{Swaroop Mishra}, \bibinfo{person}{Alisa Liu},
  \bibinfo{person}{Noah~A Smith}, \bibinfo{person}{Daniel Khashabi}, {and}
  \bibinfo{person}{Hannaneh Hajishirzi}.} \bibinfo{year}{2023}\natexlab{}.
\newblock \showarticletitle{Self-Instruct: Aligning Language Models with
  Self-Generated Instructions}. In \bibinfo{booktitle}{\emph{Annual Meeting of
  the Association for Computational Linguistics}}.
  \bibinfo{pages}{13484--13508}.
\newblock


\bibitem[Wei et~al\mbox{.}(2021)]%
        {wei2021finetuned}
\bibfield{author}{\bibinfo{person}{Jason Wei}, \bibinfo{person}{Maarten Bosma},
  \bibinfo{person}{Vincent~Y Zhao}, \bibinfo{person}{Kelvin Guu},
  \bibinfo{person}{Adams~Wei Yu}, \bibinfo{person}{Brian Lester},
  \bibinfo{person}{Nan Du}, \bibinfo{person}{Andrew~M Dai}, {and}
  \bibinfo{person}{Quoc~V Le}.} \bibinfo{year}{2021}\natexlab{}.
\newblock \showarticletitle{Finetuned language models are zero-shot learners}.
\newblock \bibinfo{journal}{\emph{arXiv preprint arXiv:2109.01652}}.
\newblock


\bibitem[Yang et~al\mbox{.}(2025)]%
        {yang2025ollama}
\bibfield{author}{\bibinfo{person}{Michael Yang}, \bibinfo{person}{Jeffrey
  Morgan}, \bibinfo{person}{Daniel Hiltgen}, \bibinfo{person}{Bruce MacDonald},
  \bibinfo{person}{Matt Williams}, \bibinfo{person}{Patrick Devine},
  \bibinfo{person}{Blake Mizerany}, \bibinfo{person}{Michael},
  \bibinfo{person}{Jesse Gross}, \bibinfo{person}{Josh},
  \bibinfo{person}{royjhan}, \bibinfo{person}{Jeremy}, \bibinfo{person}{frob},
  \bibinfo{person}{Dane Madsen}, \bibinfo{person}{Parth Sareen},
  \bibinfo{person}{Eva H}, \bibinfo{person}{Mark Ward}, \bibinfo{person}{James
  Braza}, \bibinfo{person}{Arne Müller}, \bibinfo{person}{Hernan Martinez},
  \bibinfo{person}{Ikko~Eltociear Ashimine}, \bibinfo{person}{Rapid Architect},
  \bibinfo{person}{Sam}, \bibinfo{person}{Timothy~Jaeryang Baek},
  \bibinfo{person}{slouffka}, \bibinfo{person}{Alexander~F. Rødseth},
  \bibinfo{person}{tusharhero}, \bibinfo{person}{Eli Bendersky},
  \bibinfo{person}{Lei Jitang}, {and} \bibinfo{person}{Mraiser}.}
  \bibinfo{year}{2025}\natexlab{}.
\newblock \bibinfo{booktitle}{\emph{Ollama}}.
\newblock
\urldef\tempurl%
\url{https://github.com/ollama/ollama}
\showURL{%
\tempurl}


\bibitem[Yao et~al\mbox{.}(2022)]%
        {yao2022zero}
\bibfield{author}{\bibinfo{person}{Zhewei Yao}, \bibinfo{person}{Reza~Yazdani
  Aminabadi}, \bibinfo{person}{Minjia Zhang}, \bibinfo{person}{Xiaoxia Wu},
  \bibinfo{person}{Conglong Li}, {and} \bibinfo{person}{Yuxiong He}.}
  \bibinfo{year}{2022}\natexlab{}.
\newblock \showarticletitle{{ZeroQuant}: Efficient and affordable post-training
  quantization for large-scale transformers}. In
  \bibinfo{booktitle}{\emph{Conference on Neural Information Processing
  Systems}}.
\newblock


\bibitem[Yuan et~al\mbox{.}(2023)]%
        {yuan2023no}
\bibfield{author}{\bibinfo{person}{Zhiqiang Yuan}, \bibinfo{person}{Yiling
  Lou}, \bibinfo{person}{Mingwei Liu}, \bibinfo{person}{Shiji Ding},
  \bibinfo{person}{Kaixin Wang}, \bibinfo{person}{Yixuan Chen}, {and}
  \bibinfo{person}{Xin Peng}.} \bibinfo{year}{2023}\natexlab{}.
\newblock \showarticletitle{No more manual tests? {E}valuating and improving
  {ChatGPT} for unit test generation}.
\newblock \bibinfo{journal}{\emph{arXiv preprint arXiv:2305.04207}}.
\newblock


\bibitem[Zhang et~al\mbox{.}(2025)]%
        {zhang2024exlong}
\bibfield{author}{\bibinfo{person}{Jiyang Zhang}, \bibinfo{person}{Yu Liu},
  \bibinfo{person}{Pengyu Nie}, \bibinfo{person}{Junyi~Jessy Li}, {and}
  \bibinfo{person}{Milos Gligoric}.} \bibinfo{year}{2025}\natexlab{}.
\newblock \showarticletitle{exLong: Generating exceptional behavior tests with
  large language models}. In \bibinfo{booktitle}{\emph{International Conference
  on Software Engineering}}.
\newblock


\end{thebibliography}
